%% file: main.tex
\documentclass{article}
\usepackage{comment}
\usepackage{hyperref}
\usepackage{amsmath}
\usepackage{amsthm}
\usepackage{amssymb}
\usepackage{amsfonts}
\usepackage{graphicx}
\usepackage[capitalise]{cleveref}
\usepackage{booktabs}
\usepackage{siunitx}
\usepackage{subcaption}
\usepackage{verbatim}

\usepackage{mdframed}

\hypersetup{
  colorlinks,
  citecolor=black,
  filecolor=black,
  linkcolor=black,
  urlcolor=black
}

\begin{document}

\thispagestyle{plain}
\begin{titlepage}
\begin{center}
  \large
  \textbf{Non-linear model and optimization method for a single-axis
    linear-motion energy harvester for footstep excitation}

  \vspace{0.4cm}
  \normalsize{}
  Michael N. Struwig, Riaan Wolhuter, Thomas Niesler

  \vspace{0.9cm}
  \textbf{Abstract}
\end{center}
We propose and develop an electrical and mechanical system model of a
single-axis linear-motion kinetic energy harvester for impulsive excitation that
allows its generated load power to be numerically optimised as a function of
design parameters. The device consists of an assembly of one or more spaced
magnets suspended by a magnetic spring and passing through one or more coils
when motion is experienced along the axis. The design parameters that can be
optimised include the number of coils, the coil height, coil spacing, the number
of magnets, the magnet spacing and the physical size. We use the proposed model
to design optimal energy harvesters for the case of impulse-like motion like
that experienced when attached to the leg of a human. We generate several
optimised designs, ranked in terms of their predicted load power output. The
three best designs are subsequently constructed and subjected to controlled
practical evaluation while attached to the leg of a human subject. The results
show that the ranking of the measured output power corresponds to the ranking
predicted by the optimisation, and that the numerical model correctly Predicts
the relative differences in generated power for complex motion. It is also found
that all three designs far outperform a baseline design. The best energy
harvesters generated an average power of 3.01mW into a 40$\Omega$ test load
when driven by footsteps whose measured peak impact was approximately 2.2g. With
respect to the device dimensions, this corresponds to a power density of
179.380$\mu\text{W/cm}^3$.
\\

\\

\noindent
Declaration of interest: none.

\end{titlepage}

\input{introduction}
\input{architecture}
\input{analyticalModels}
\input{designApplication}
\input{practicalTesting}
\input{applications}

\input{conclusion}
\input{acknowledgements}
\input{funding}
\newpage
\bibliographystyle{IEEEtran} \bibliography{Masters}
\end{document}

%% file: introduction.tex
\section{Introduction} \label{sec:introduction} Energy harvesting has become an
increasingly popular field as researchers and industry alike attempt to discover
and improve ways of powering electrical devices in situations where conventional
sources of power are unavailable \cite{Conrad2008} \cite{Vullers2009}
\cite{Sudevalayam2011}. One source is kinetic energy, where the mechanism of
electromagnetic induction is used to generate electrical energy
\cite{Gilbert2008}. The literature has described a wide variety of
electromagnetic kinetic microgenerators, many of which are modelled as linear
spring-damper systems that harvest energy from sources of harmonic vibration
\cite{Bobryk2016} \cite{Wang2017} \cite{Yang2017}. More recently, increasingly
complex non-linear models have been proposed in a quest to harvest energy from a
wider variety of kinetic sources \cite{Haroun2015} \cite{Kecik2017}
\cite{Marszal2017}. However, for both linear and non-linear models, the analysis
based on electromagnetic first principles results in a significant degree of
mathematical complexity and requires the use of parameters that are difficult to
determine \cite{SoaresdosSantos2016}. This has hindered the development of
techniques that allow the parametric optimization of kinetic energy harvesting
beyond resonant-frequency optimization and resistive load matching
\cite{Berdy2015} \cite{Serre2008}. When parametric optimization is performed,
it is typically in the form of experimental iteration \cite{Kwon2013}
\cite{Wickenheiser2010}, or considers a very limited number of parameters
\cite{Ylli2015} \cite{Saravia2017}.

In contrast to harmonic vibration, very little attention has
been given to energy harvesters driven by impulse-like accelerations
\cite{Carroll2012}. This currently severely limits the design of a kinetic
energy harvesting device for an environment in which the primary source of
energy is impulse-like acceleration, such as that resulting from the footstep of
a person or animal. Methods developed for harmonic vibration
cannot accurately be applied in these situations.

The initial motivation for the work we present here was to enable the design of
self-powering animal-borne sensors for use in the monitoring and conservation of
large wildlife. In this situation, the source of energy is the impulse-like
motion of the animal's leg, while severe size and weight limitations make it
essential to maximize the harvested energy. Despite major advancement in the
sophistication of animal-tracking collars, such as the inclusion of GPS,
on-board data transmission and on-board behaviour
classification \cite{leRoux2017}, battery life remains the greatest limitation
\cite{Blackie2010} \cite{hebblewhite2007}.

We propose a non-linear mechanical model for a single-axis electromagnetic
kinetic microgenerator that allows the constrained parametric optimization of
design parameters in order to maximize the average power supplied to the
attached load. This is achieved by developing an electrical model for a
simplified configuration of the energy harvesting device for a form of
non-harmonic motion. This model is applied to the analysis of more complex
configurations, thereby providing a means of optimizing the microgenerator
design. It should be noted that despite sharing similar architecture features,
the proposed model and optimization methodology differs significantly from those
typically found in literature \cite{Masoumi2016} due to the focus on harvesting
energy from non-vibrational sources. Additionally, our approximate analytical
model is in contrast to the use of first-principle electromagnetic techniques
previously proposed \cite{Donoso2009}\cite{VonBuren2007} and offers the
advantage of much greater computational speed, once defined, in comparison to
performing a simulation with finite element analysis (FEA) due to the proposed
model's parametric nature. We demonstrate the effectiveness of our method by
using it to determine a number of optimal microgenerator designs, and
subsequently constructing and practically evaluating these for footstep input.

%% file: architecture.tex
\section{Microgenerator architecture}
\begin{figure}[ht]
  \centering \includegraphics[width=0.8\textwidth]{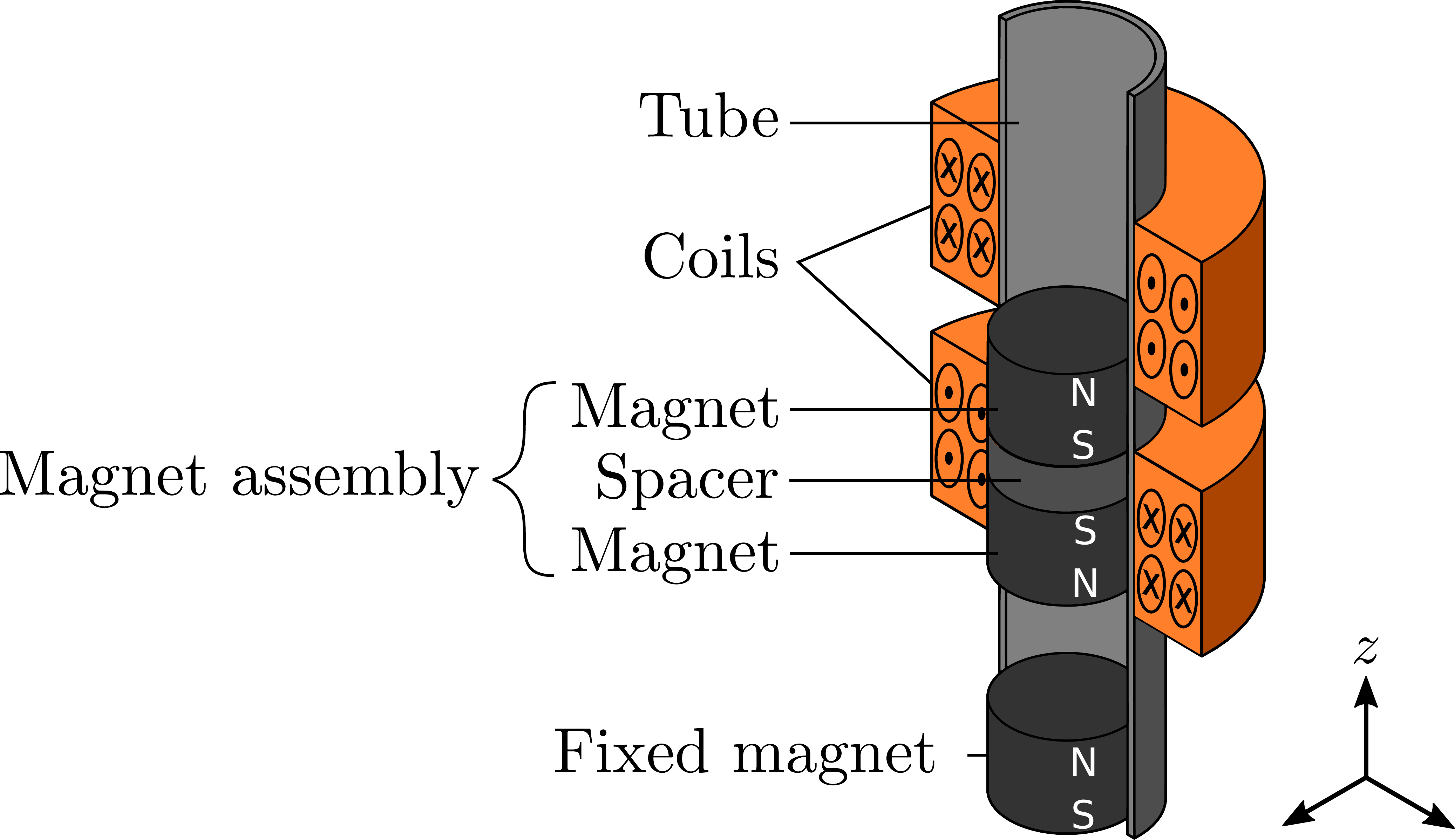}
  \caption{Basic structure of the linear kinetic energy harvester, showing
    multiple coils and multiple magnets. The magnets and coils are pole-matched.
    A single fixed magnet at the base provides a magnetic spring.}
  \label{fig:basic_design}
\end{figure}

The kinetic energy harvester we consider consists of a hollow circular tube,
inside of which a set of one or more magnets is able to move, as illustrated in
\cref{fig:basic_design}. The primary axis is parallel to the limb of the person
or animal. A number of uniformly spaced
coils are wound around the outside of the tube. The magnet assembly consists of
a number of permanent magnets that are arranged with alternating polarity and
separated by a spacer of some ferrous material. A fixed pole-matched magnet is
placed at the bottom of the tube. This non-linear magnetic spring allows the
magnet assembly to oscillate and for gravity to reset its position after motion.
The non-linear characteristic of the magnet spring increases model complexity,
but affords several benefits. Firstly, the non-linear spring
can push the magnet assembly upwards, but cannot pull it downwards, thereby
increasing the range of motion of the magnet assembly. Secondly,
the magnetic spring is more consistent and less prone to mechanical failure than
a mechanical spring, which is of critical importance for our eventual intended
application in wildlife monitoring. Finally, it is simpler to assemble, as the
magnetic spring consists of the same type of magnet used in the magnet assembly,
and does not require manufacture or sourcing of a specialized mechanical spring.

Once selected, the properties of the individual permanent magnets, copper wire
and winding density are considered to be fixed. The remaining variables, which
include the footstep forces, coil height, electrical load, number of magnets,
number of coils and various coil properties are varied to optimize the design of
the kinetic energy harvester.

%% file: analyticalModels.tex
\section{Analytical models for microgenerator optimization} \label{sec:model}
This section develops an analytical model for the kinetic energy harvester. The
model consists of a mechanical system model, describing the mechanical responses
of the kinetic energy harvester, the footstep model that drives the mechanical
model and an electrical system model that describes the electrical output.

\subsection{Footstep model}

\begin{figure}[ht]
  \centering \includegraphics[width=0.8\textwidth]{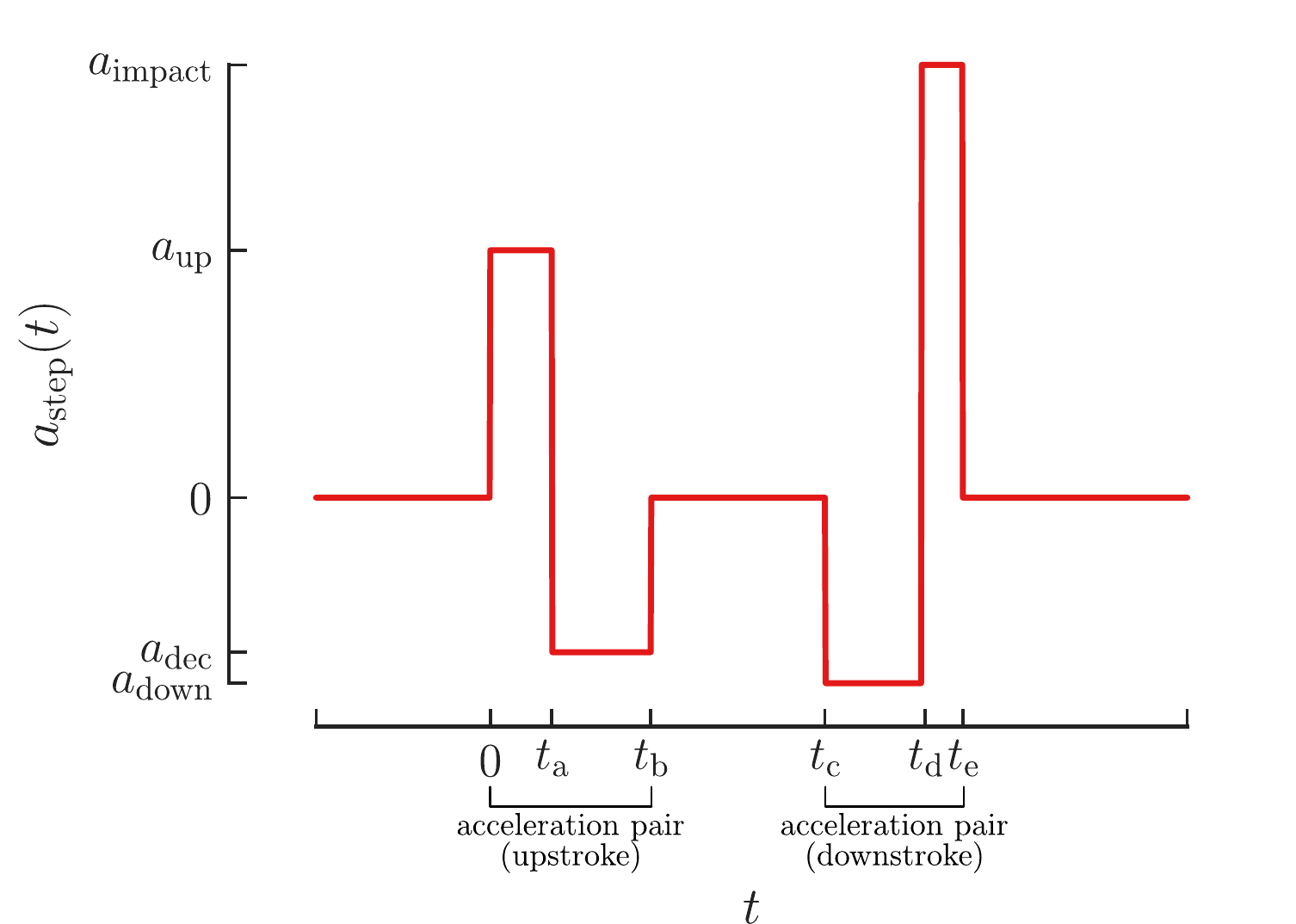}
  \caption{The footstep is modelled as two acceleration pairs; one for the
    upstroke and another for the downstroke. The two
    stroke pairs are separated by the time interval $t_{\text{b}} < t \leq
    t_{\text{c}}$. \label{fig:footstepSymbols} }
\end{figure}

\begin{figure}[ht]
  \centering \includegraphics[width=0.8\textwidth]{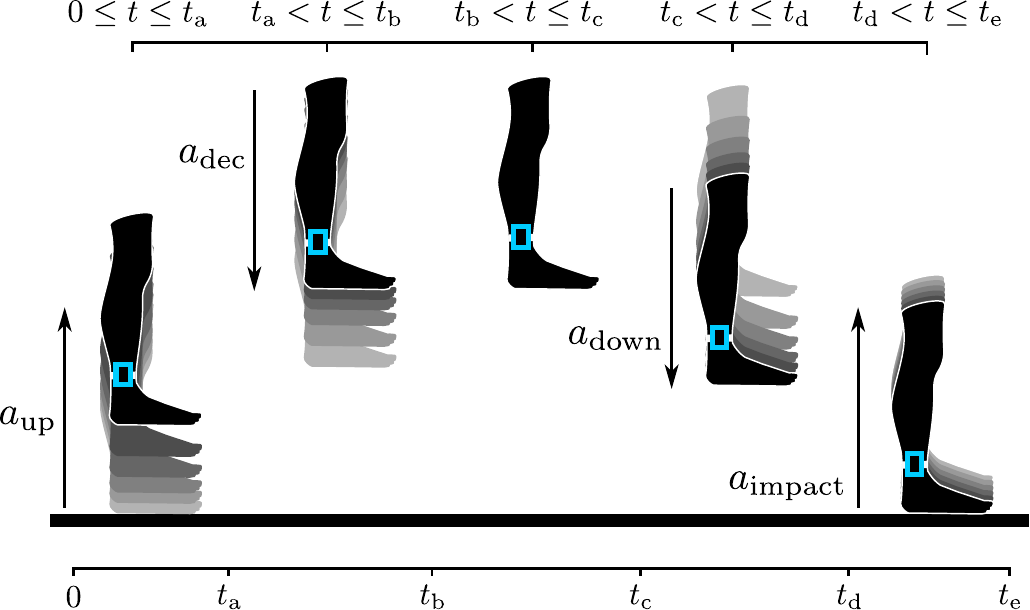}
  \caption{Shows the motion of the leg at different stages during the
    footstep cycle and the resulting direction of acceleration experience by the
    kinetic microgenerator body. \label{fig:footstep_extra} }
\end{figure}

The acceleration $a_{\text{step}}$ of the footstep is modelled as a piece-wise
constant non-periodic function, given by \cref{eq:f_step} and shown graphically
in \cref{fig:footstepSymbols}. This model is based on a simplified footstep
cycle using measurements taken from an accelerometer attached to the leg of a
human when walking and aligned with corresponding video footage. The constant
accelerations $a_{\text{up}}$, $a_{\text{dec}}$, $a_{\text{down}}$ and
$a_{\text{impact}}$ respectively represent the average acceleration experienced
by the microgenerator body during the upstroke, deceleration, downstroke and
impact phases of a single footstep cycle as shown in \cref{fig:footstep_extra}.

\begin{equation}
  a_{\text{step}}(t) =
  \begin{cases}
    a_{\text{up}}, & \text{for}\ 0 \leq t \leq t_{\text{a}}\\
    a_{\text{dec}}, & \text{for}\ t_{\text{a}} < t \leq t_{\text{b}} \\
    0, & \text{for}\ t_{\text{b}} < t \leq t_{\text{c}}\\
    a_{\text{down}}, & \text{for}\ t_{\text{c}} \leq t \leq t_{\text{d}} \\
    a_{\text{impact}}, & \text{for}\ t_{\text{d}} < t \leq t_{\text{e}} \\
    0, & \text{for}\ t_{\text{e}} < t,
  \end{cases}
  \label{eq:f_step}
\end{equation}

Assuming the foot to be motionless in the $z$-direction at $t=0$ and again at
$t=t_{\text{b}}$, $t_{\text{a}}$ and $t_{\text{b}}$ can be calculated from the
kinematic equations of motion.

\begin{align}
  t_{\text{a}} &= \sqrt{\frac{2a_{\text{dec}}s_{\text{h}}}{a_{\text{up}}(a_{\text{dec}}-a_{\text{up}})}} \label{eq:time_a}\\
  t_{\text{b}} &= \frac{2s_{\text{h}}}{a_{\text{up}}t_{\text{a}}}\left(1+\frac{a_{\text{dec}}}{a_{\text{up}}-a_{\text{dec}}}\right) + t_{\text{a}}. \label{eq:time_b}
\end{align}

Assuming further that the foot is motionless in the $z$-direction at
$t=t_{\text{c}}$ and again at $t=t_{\text{e}}$, analogous equations can be
written for $t_{\text{d}}$ and $t_{\text{e}}$ in terms of $a_{\text{down}}$ and
$a_{\text{impact}}$. In this way the values of $a_{\text{up}}, a_{\text{dec}},
a_{\text{down}}$ and $a_\text{impact}$ as well as the maximum vertical
displacement $s_\text{h}$ can be selected to match any practically measurable
footstep-like motion.

The piecewise-constant acceleration approximation shown
in~\cref{fig:footstepSymbols} was adopted for two reasons. Firstly, the utilized
accelerometer data was sampled at 40Hz, which is too sparse to use directly when
computing a numerical solution for \cref{eq:mech_system}. Secondly, by selecting
the acceleration values of each phase of the footstep to match measured
accelerometer data, $a_{\text{step}}(t)$ can approximate a wide variety of
footstep-like motion, which potentially opens up the device's application in
other fields, such as energy harvesting on wildlife.

\subsection{Mechanical system model}
\begin{figure}
  \centering \includegraphics[width=0.50\textwidth]{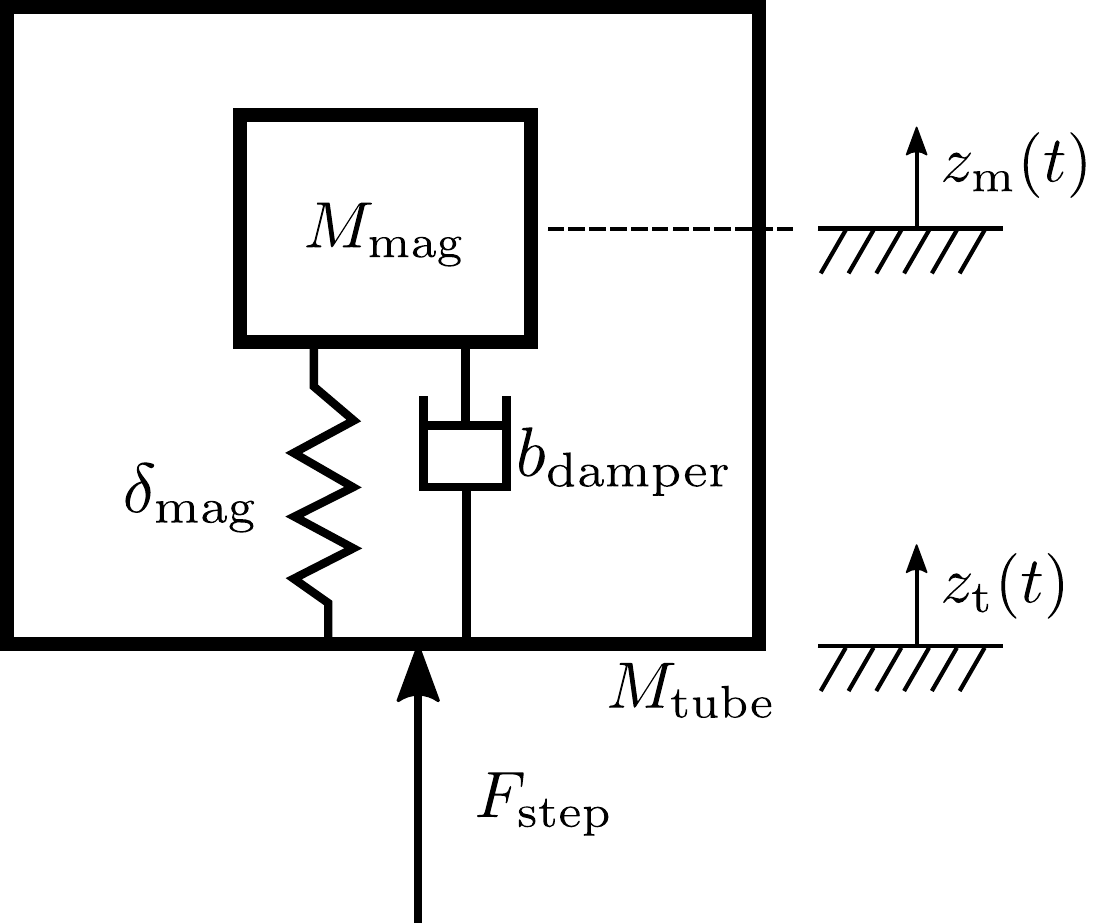}
  \caption{Idealized mechanical model of microgenerator. The mass of the magnet
    assembly $M_{\text{mag}}$ is suspended by the nonlinear magnetic spring
    $\delta_{\text{mag}}$ and experiences mechanical losses through friction
    with constant $b_{\text{damper}}$ through the walls of the tube.}
  \label{fig:mech_system}
\end{figure}

The device in \cref{fig:basic_design} is modelled by a mechanical system
consisting of a mass $M_{\text{mag}}$ representing the magnet assembly attached
to the bottom of the outer tube via a non-linear magnetic spring with force
$\delta_{\text{mag}}$ and a damper representing energy losses that are
proportional to the relative velocity between the magnet assembly and the tube
with constant $b_{\text{damper}}$. Similar models have been used to design
energy harvesters, but these employ a linear spring $\delta_{\text{spring}}$ to
facilitate the oscillatory motion of the magnet assembly \cite{Ylli2015}
\cite{Khan2014} \cite{Zeng2013}. In contrast, our design makes use of a magnetic
spring $\delta_{\text{mag}}$. This differs from other work in the literature
that utilize a magnetic spring \cite{SoaresdosSantos2016} \cite{Masoumi2016} ,
as we utilize a single repelling magnet at the bottom of the device, as shown in
\cref{fig:basic_design}. The resulting mechanical model of the system is shown in
\cref{fig:mech_system}.

The value $z_{\text{m}}(t)$ represents the position of the magnet assembly
relative to the bottom of the magnet in the assembly and $z_{\text{t}}(t)$
represents the position of the microgenerator tube relative to the top of the
fixed magnet. The magnetic spring force $\delta_{\text{mag}}$ is a non-linear
function of the relative displacement between the magnet assembly and the tube.
The mechanical system can be reduced to a set of first-order differential
equations by defining $x_1 = z_{\text{m}}$, $x_2 = \dot{z_{\text{m}}}$, $x_3 =
z_{\text{t}}$, $x_4 = \dot{z_{\text{t}}}$, to find:
\begin{equation}
  \begin{bmatrix}
    \dot{x_1} \\
    \dot{x_2} \\
    \dot{x_3} \\
    \dot{x_4} \\
  \end{bmatrix}
  =
  \begin{bmatrix}
    x_2 \\
    \big(\delta_{\text{mag}}(x_1-x_3)+b_{\text{damper}}\cdot (x_2-x_4)\big)/M_{\text{mag}} \\
    x_4 \\
    a_{\text{step}}(t)
  \end{bmatrix}.
  \label{eq:mech_system}
\end{equation}

To solve \cref{eq:mech_system}, the continuous function $\delta_{\text{mag}}$
and the function $a_{\text{step}}$ must be known. The force between two
permanent magnets as a function of the distance between them is difficult to
determine analytically because it requires a knowledge of quantities that are
extremely difficult to describe, such as the $B-$field vector of magnetic flux
density, the vector of the magnetic dipole moment $m$, and the
interactions of these properties between the two magnets \cite{Vokoun2009}.

We propose a simple approximation of $\delta_{\text{mag}}(z)$ that can be
obtained by considering Coulomb's Law, which can be used to describe the force
between two hypothetical magnetic monopoles of strength $m_1$ and $m_2$ as
inversely proportional to the squared distance $r$ between them.

\begin{align}
  \delta_{\text{mag}}(z) &= \frac{\mu_{0}m_1 m_2}{4 \pi z^2}
                           \label{eq:coulombs_law}
\end{align}

For comparison, we consider a power series model that is commonly utilized
in literature to model the force between two magnets \cite{Mann2009}:

\begin{equation}
  \delta_{\text{mag}}(z) = \sum_{n=0}^{3} a_{n}z^{n}.
  \label{eq:mag_power_series}
\end{equation}

\begin{figure}[ht]
  \centering \includegraphics[width=0.8\textwidth]{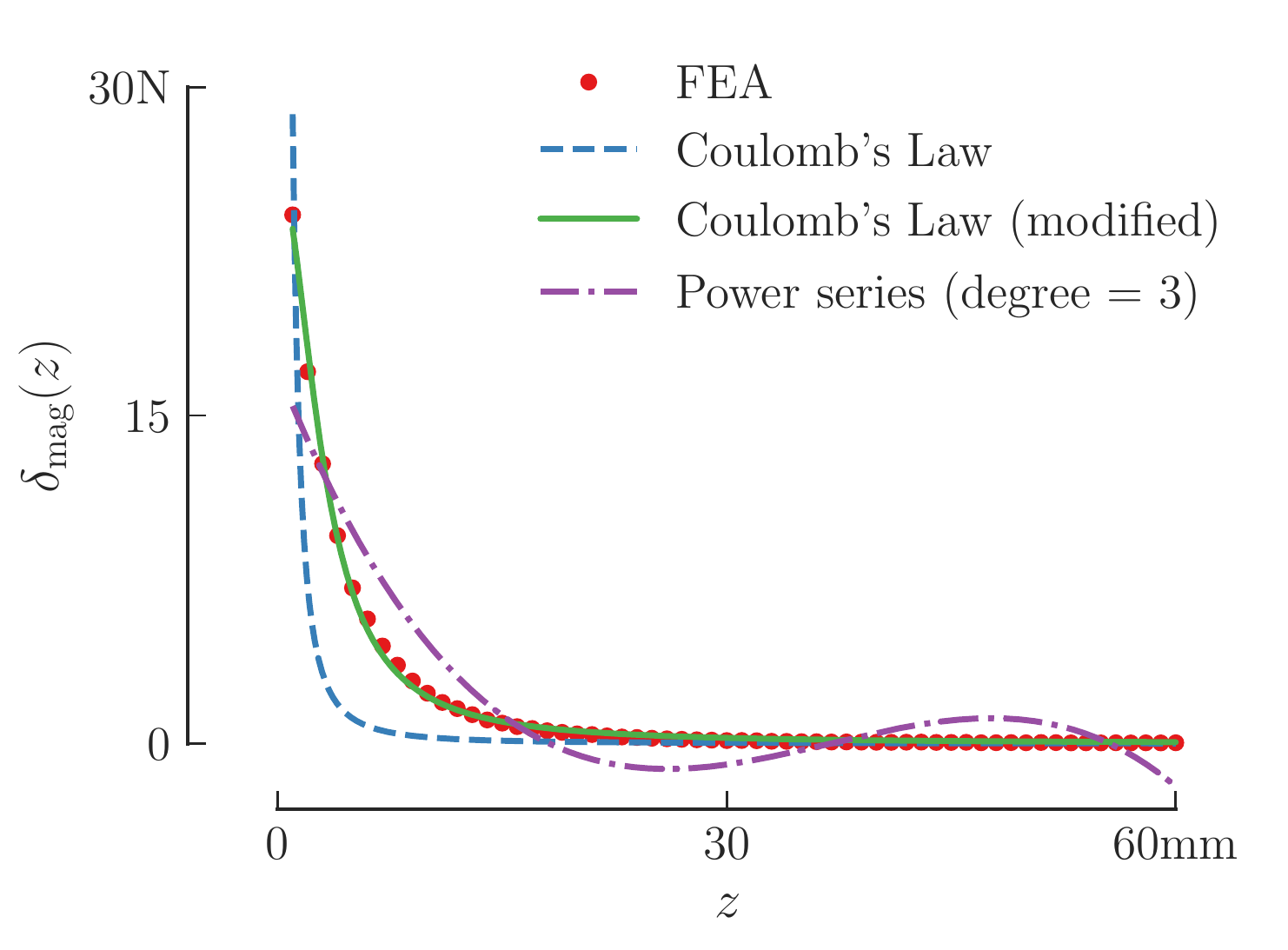}
  \caption{The repelling force between two pole-matched cylindrical N35-grade
    NdFeB magnets with height of 10mm and a radius of 5mm. The power series
    model provides a generally poor approximation for the true force between two
    magnets at all distances. In contrast, Coulomb's Law provides a poor
    approximation when the two magnets are close to one another, yet is accurate
    when they are far apart. By including the additional parameter $G$ in the
    denominator of Coulomb's Law, a much better correspondence with FEA is
    achieved for smaller separation distances
    $z$. \label{fig:magSpringCombined}}
\end{figure}

Discrete values of $\delta_{\text{mag}}(z)$ can be obtained by simulation using
FEA analysis of two identical cylindrical magnets aligned vertically along their
N-S axes such that their poles are matched. The force experienced by the second
magnet can then be obtained using FEA for a discrete set of separating distances
$z$. The results of such a simulation are compared with the approximation given
by \cref{eq:coulombs_law,eq:mag_power_series} in \cref{fig:magSpringCombined}
for constants of $m_1=m_2=4.119\text{A} \cdot \text{m}$ in the case of
\cref{eq:coulombs_law} and for coefficients $a_o =
\num{1.709e1},~a_1=\num{-1.713e03},~a_2=\num{5.082e04}~\text{and}~a_3=\num{-4.594e05}$
in the case of \cref{eq:mag_power_series}. The constants and coefficients are
found using a typical least-squares curve-fit procedure.

We see that the power series approximation gives a poor approximation of the
force between two magnets in our case. We also see that, while the approximation
is poor when the magnets are close together, the Coulomb's
Law approximation is good when the magnets are far apart.

To improve the approximation given by Coulomb's Law, we modify \cref{eq:coulombs_law} as shown in
\cref{eq:coulombs_modified}, creating a modified version of Coulomb's Law:

\begin{equation}
  \delta_{\text{mag}}(z) = \frac{\mu_0 m_1 m_2}{4 \pi z^2 + G}.
  \label{eq:coulombs_modified} 
\end{equation}

The constant parameter $G$ sharpens the knee of the original Coulomb's Law curve
for small values of $z$, and is determined by fitting
\cref{eq:coulombs_modified} to values of $\delta_{\text{mag}}$ obtained by
numerical simulation, giving $m_1=m_2=15.302\text{A} \cdot {m}$ and $G=\num{1.125e-4}$. This
provides a much better approximation of the true value of $\delta_{\text{mag}}$
in a closed form, shown in \cref{fig:magSpringCombined} , which will allow FEA
to be sidestepped later.

\subsection{Electrical system model}
The primary goal of the electrical system model is to describe the power
delivered to the load, parameterized in terms of a set of important design
parameters. A model will first be developed to describe the
idealized movement of a single magnet passing through a single coil at a
constant velocity. Subsequently, this will be extended to include configurations
with multiple coils and multiple magnets for the same type of movement. Finally,
it will be demonstrated that the estimated power produced with this simple
motion is representative of the true power that produced when the device is
operated practically in the field.

\begin{figure}[ht]
  \centering
  \includegraphics[width=0.4\textwidth]{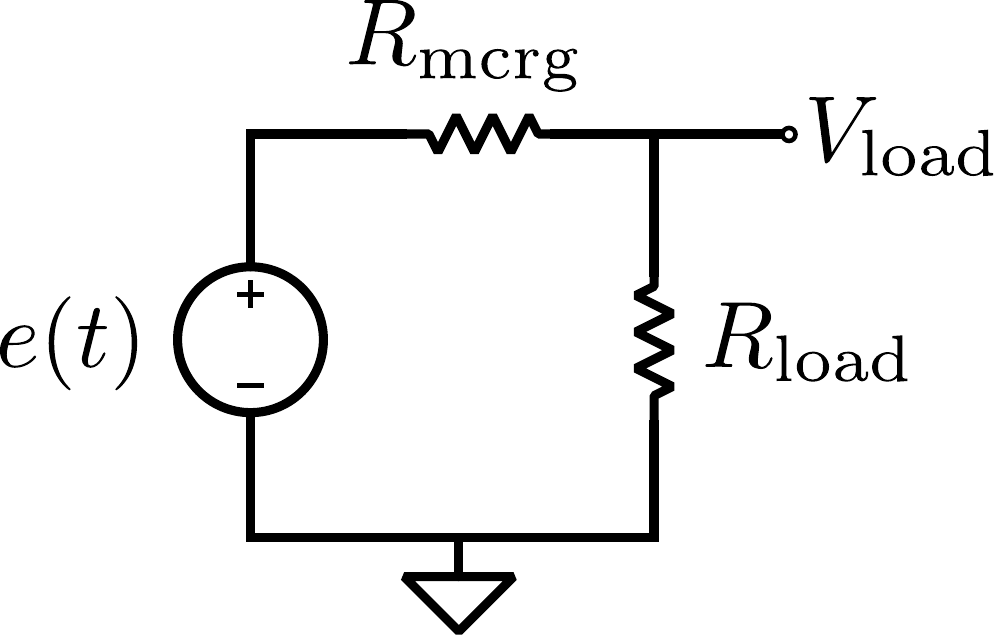}
  \caption{\label{fig:basic_circuit} The kinetic microgenerator powering a load
    is modelled as a voltage source and two resistors connected in series. The
    first resistor, $R_{\text{mcrg}}$, represents the internal resistance of the
    kinetic microgenerator and the second, $R_{\text{load}}$, represents the
    load. The voltage $e(t)$ represents the instantaneous open-circuit EMF
    produced by the kinetic microgenerator in volts.}
\end{figure}

A resistive load $R_{\text{load}}$ in series with a kinetic microgenerator with
internal resistance $R_{\text{mcrg}}$, as shown in \cref{fig:basic_circuit},
dissipates an instantaneous power $P_{\text{load}}(t)$ in the load as given by
\begin{equation}
  P_{\text{load}}(t) = V_{\text{load}}(t)^2/R_{\text{load}} \label{eq:p_load}
\end{equation}

where it can be shown that
\begin{equation}
  V_{\text{load}}(t) = e(t) \frac{R_{\text{load}}}{R_{\text{load}}+R_{\text{mcrg}}} \label{eq:v_load}
\end{equation}

and where $e(t)$ is the instantaneous open-circuit EMF produced by the kinetic
microgenerator, in volts. The RMS power $\bar{P}_{\text{load}}$ delivered to the
load is then given by:

\begin{equation}
  \bar{P}_{\text{load}} = \frac{e^2_{\text{RMS}} R_{\text{load}}}{(R_{\text{mcrg}} + R_{\text{load}})^2}.
  \label{eq:power_avg}
\end{equation}
Hence, by modeling $e(t)$ and $R_{\text{mcrg}}$, $P_{\text{load}}$ and
$\bar{P}_{\text{load}}$ can be determined.

\subsubsection{Single coil, single magnet configuration} \label{sec:scsm} We
first consider the case of a kinetic microgenerator consisting of a single coil
and a single magnet, before extending the analysis to a more general case.

\begin{figure}[ht]
  \centering \includegraphics[width=0.5\textwidth]{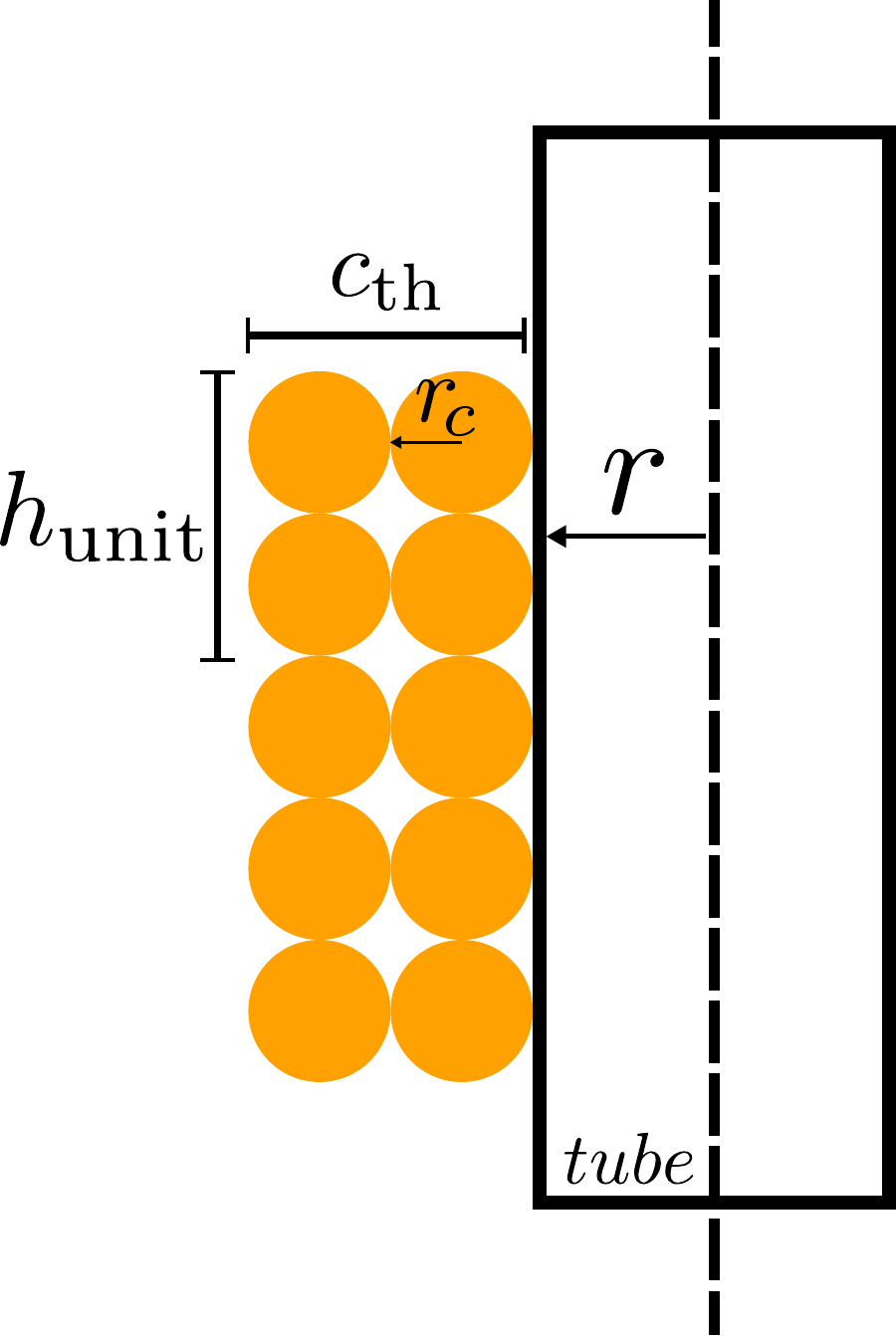}
  \caption{Cross section of microgenerator body and coil with square tiling of
    turns. The coil consists of $N$ number of packed turns, with wire radius
    $r_{\text{c}}$, creating a coil with thickness $c_{\text{th}}$ and height
    $h$. \label{fig:square_coil}}
\end{figure}

\Cref{fig:square_coil} shows a coil defined by three primary parameters; the
turn density $\gamma$ (measured in turns per mm), the height of the coil $h$
(measured in mm) and the resistance per turn $\beta$ (measured in Ohms per
turn). The turn density is given by

\begin{align}
  \gamma &= \frac{N}{h_{\text{unit}}} \nonumber\\
         &= f_{\text{f}}\frac{c_{\text{th}}}{\pi r^2_{\text{c}}},
           \label{eq:turn_density}
\end{align}
where $N$ is the number of packed coil turns, $f_\text{f} = 0.7$ is the fill factor ratio of the turns, $c_{\text{th}}$
is the coil thickness in mm and $r_{\text{c}}$ is the radius of the copper wire
in mm.

Assuming square tiling of the turns, the coil resistance $R_{\text{coil}}$ is
given by
\begin{equation}
  R_{\text{coil}} = N \pi R_{\text{gauge}}(2r+2r_{\text{c}}+c_{\text{th}}),
  \label{eq:r_coil}
\end{equation}
where $R_{\text{gauge}}$ is the resistance per unit length of the copper wire in
\si{\ohm\per\mm}. From \cref{eq:r_coil}, the resistance per turn $\beta$ in mm
is given by:

\begin{align}
  \beta &= \frac{R_{\text{coil}}}{N} \nonumber \\
        &= \pi R_{\text{gauge}}(2r + 2r_{\text{c}} + c_{\text{th}}).
          \label{eq:ohm_per_turn}
\end{align}
Combining \cref{eq:turn_density} and \cref{eq:ohm_per_turn} allows the coil
resistance to be expressed in terms of all coil parameters:

\begin{equation}
  R_{\text{coil}} = \beta \gamma h.
  \label{eq:r_coil_param}
\end{equation}

With the coil model defined, we consider the effect of the coil height $h$ on
the open-circuit EMF $e(t)$. This is achieved by varying $h$ while keeping all
other coil parameters constant. With the magnetic field vector
$\vec{\boldsymbol{B}}$ known, an expression for $e(t)$ can be derived using the
concept of motional EMF,

\begin{equation}
  e(t) = \int_c (\vec{\boldsymbol{u}} \times \vec{\boldsymbol{B}}) \cdot \vec{dl}_c,
  \label{eq:motional_emf}
\end{equation}
where $\vec{\boldsymbol{u}}$ is the velocity of a conductor moving through
magnetic field $\vec{\boldsymbol{B}}$ with elemental length $\vec{dl}$ and where
$c$ indicates the path of integration. Alternatively, Faraday's law of induction
can be used to derive an expression for the EMF.

\begin{equation}
  e(t) = -\frac{d}{dt} \int_s \vec{\boldsymbol{B}} \cdot d\vec{\boldsymbol{s}}, \label{eq:faraday_law}
\end{equation}
where $d\vec{\boldsymbol{s}}$ is the infinitesimal surface element of a surface
$s$ enclosed by the conductor. However, the analytic solutions of
\cref{eq:motional_emf,eq:faraday_law} for the problem at hand are prohibitively
complex. Hence the characteristics of $e(t)$ will be investigated numerically
using FEA.

\begin{figure}[ht]
  \centering \includegraphics[width=0.5\textwidth]{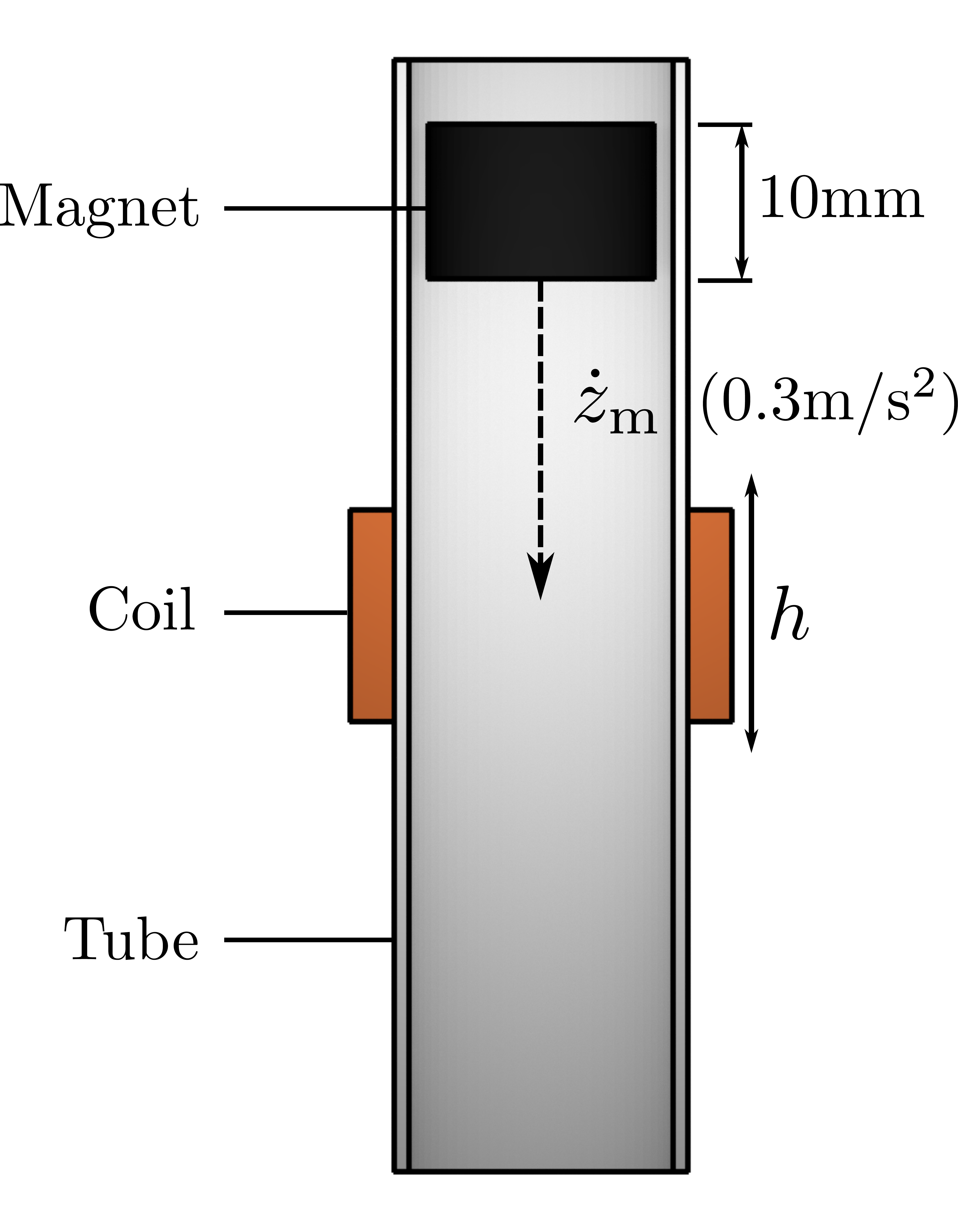}
  \caption{The FEA model used to determine the effect of $h$ on the open-circuit
    EMF $e(t)$ of a single coil, single magnet kinetic
    microgenerator. \label{fig:1c1mdiagram} }
\end{figure}

\begin{figure}[ht]
  \centering \includegraphics[width=0.8\textwidth]{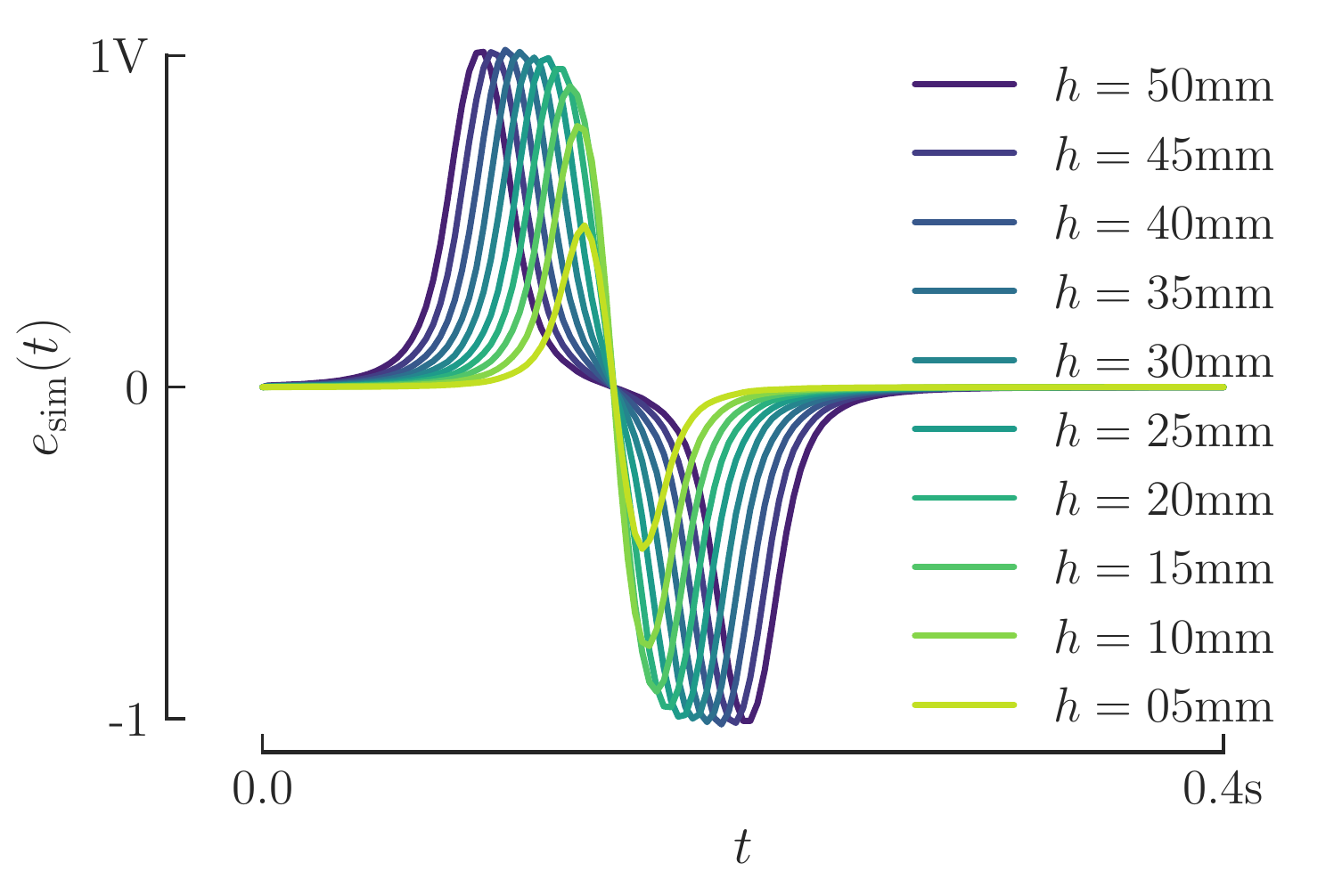}
  \caption{FEA of the open-circuit EMF produced by a single-coil, single magnet design for
    varied $h$ when $\dot{z}_{\text{m}} = 0.3\text{m/s}^2$. \label{fig:oc_emf} }
\end{figure}

To do this, we consider the idealized simple transient model shown in
\cref{fig:1c1mdiagram}. The model consists of a single-magnet, single-coil
microgenerator, in which the magnet is assumed to pass through the coil with a
constant velocity. \Cref{fig:oc_emf} shows the result of numerical simulations,
by means of FEA, for a range of values of $h$, while a constant velocity
$\dot{z}_{\text{m}}=0.3\text{m/s}^2$ is used as well as constant values for the
parameters $c_{\text{th}}, \gamma, \beta$ and the magnet properties. We see that
the induced EMF consists of a pulse of positive amplitude followed by one of
negative amplitude. These positive and negative pulses occur as the magnet
enters and exits the coil, respectively. The zero-crossing occurs at the instant
the magnet is at the centre of the coil and the rate of change of flux becomes
zero.

\begin{figure}[ht]
  \centering \includegraphics[width=0.8\textwidth]{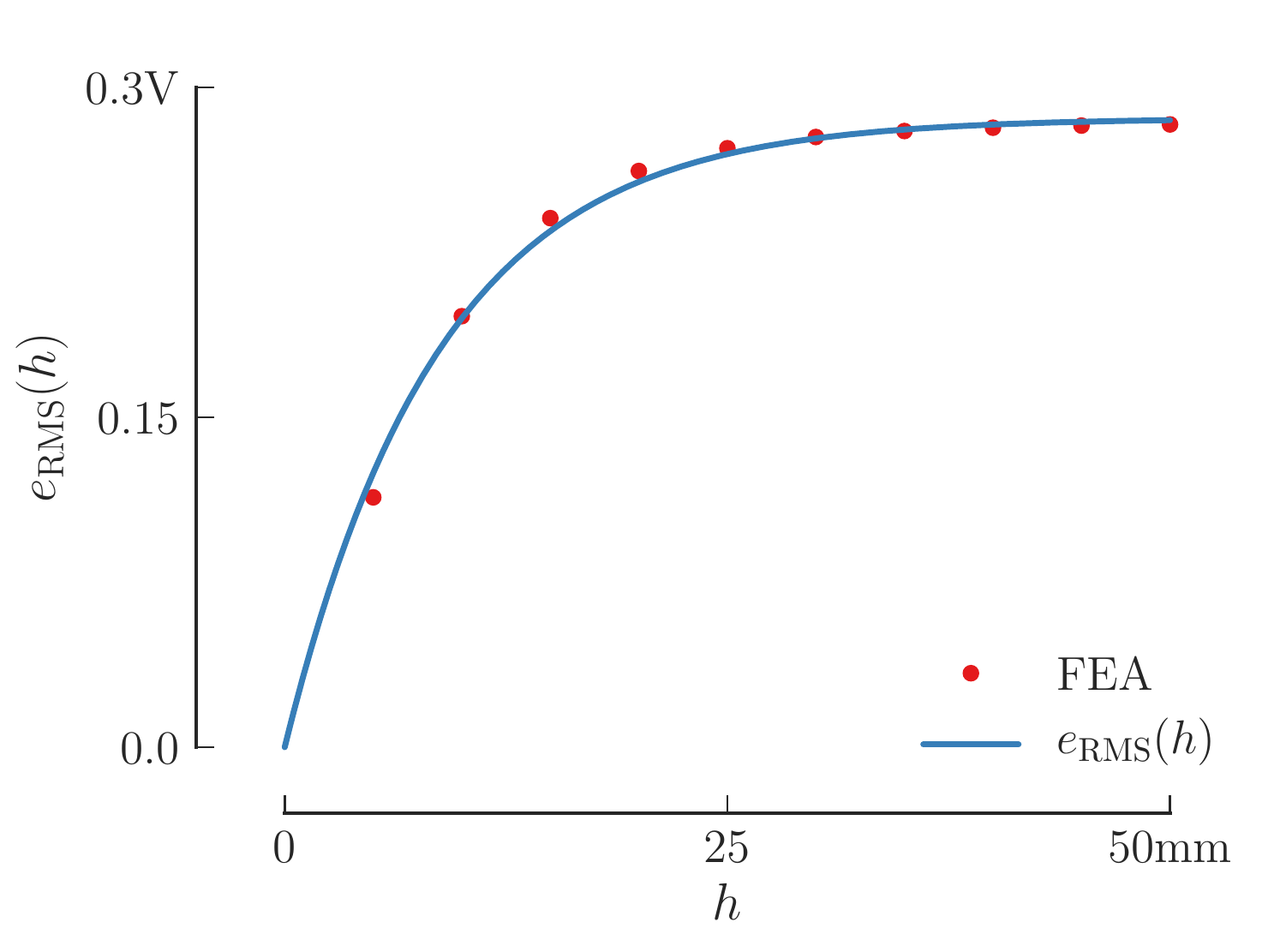}
  \caption{\Cref{eq:oc_rms} approximates the open-circuit EMF, as a function of
    the coil height, produced by a single-coil, single-magnet
    microgenerator.\label{fig:oc_rms} }
\end{figure}

It can be seen that both the shape and the peak amplitudes of the waveform
change substantially with $h$, making it difficult to infer the effect of $h$ on
$e(t)$ directly. Instead we consider the RMS of the FEA waveforms shown in
\cref{fig:oc_emf}. It was found empirically that this RMS voltage can be
approximated by \cref{eq:oc_rms}, with the constants $A$ and $\alpha$ found from
the FEA for the values of $h$ in question. A comparison of the RMS voltage
values obtained by FEA and the $e_{\text{RMS}}(h)$ estimated using
\cref{eq:oc_rms} is shown in \cref{fig:oc_rms}.

\begin{equation}
  e_{\text{RMS}}(h)  = A \cdot (1-\exp{(-\alpha h)}).
  \label{eq:oc_rms}
\end{equation}

For a given set of coil parameters $c_{\text{th}}$, $\gamma$, $\beta$ and a
given magnetic field, the results in \cref{fig:oc_rms} show that the
open-circuit RMS EMF rapidly increases with $h$ before approaching a constant
value. This indicates that the coil height $h$ strongly influences the amount of
harvestable energy up to a point, after which further increases provide ever
diminishing returns.

\begin{figure}[ht]
  \centering \includegraphics[width=0.8\textwidth]{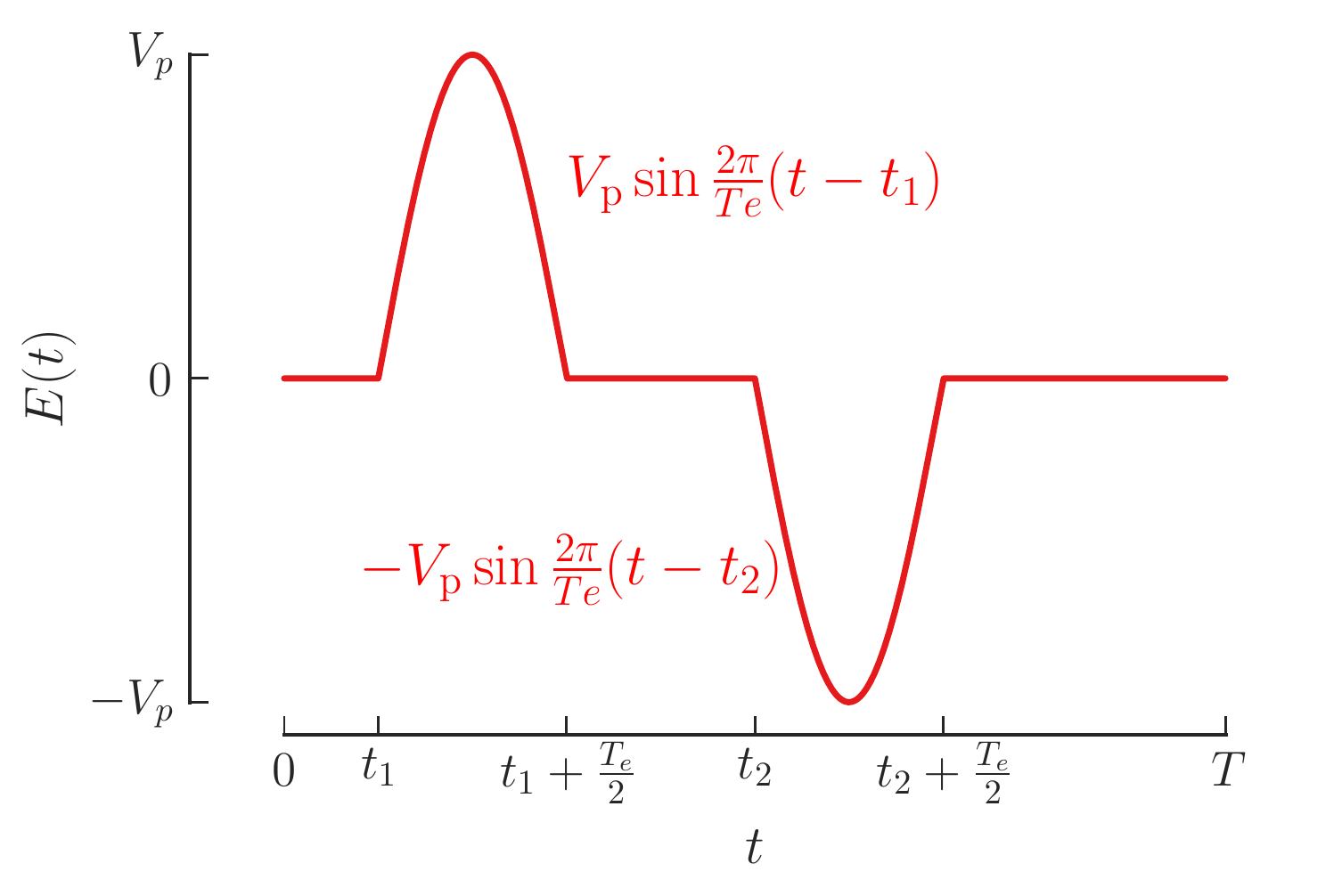}
  \caption{The function $E(t)$ consists of two half-sine lobes, given by
    \cref{eq:e1_sin}, that occur somewhere between $0 \leq t \leq
    T$.\label{fig:sin_within_global} }
\end{figure}

As a next step, we approximate the waveforms shown in \cref{fig:oc_emf} by the
superposition $E(t)$ of two half periods of a sinusoid with period $T_e$, as shown
in~\cref{fig:sin_within_global} and described by \cref{eq:e1_sin}.
\begin{equation}
  E(t) =
  \begin{cases}
    V_{\text{p}} \sin{\frac{2\pi}{T_e} (t-t_1)}, & \text{for } t_1 \leq t \leq t_1 + T_e/2 \\
    -V_{\text{p}} \sin{\frac{2\pi}{T_e}(t-t_2)}, & \text{for } t_2 \leq t \leq t_2 + T_e/2\\
    0, & \text{otherwise}.
  \end{cases}
  \label{eq:e1_sin}
\end{equation}

The positive half-period in \cref{eq:e1_sin}, beginning at $t_1$, approximates the EMF
generated as the magnet enters the coil, and the negative half-period, beginning at
$t_2$, approximates the EMF generated as the magnet exits the coil.
\Cref{eq:e1_sin} is zero when the rate of change of flux in the coil is zero.
This occurs when the magnet is not in close proximity to the coil (before $t =
t_1$ and after $t=t_2 + T_e / 2$) or when the magnet is approximately centered
within the coil at $t_1 \leq t \leq t_2 + T_e / 2$.

The RMS of $E(t)$ over a period $T$ can be expressed as:

\begin{equation}
  e_{\text{RMS}}(V_{\text{p}}, T_e, T)  = \sqrt{\frac{T_{e}}{2T}} V_{\text{p}}, \label{eq:RMS_single_sine}
\end{equation}
                                        
The constants $V_{\text{p}}$ and $T_e$ can be determined from the open-circuit
EMF waveforms determined by FEA, such as those shown in~\cref{fig:oc_emf}.

\begin{figure}[ht]
  \centering
  \includegraphics[width=0.8\textwidth]{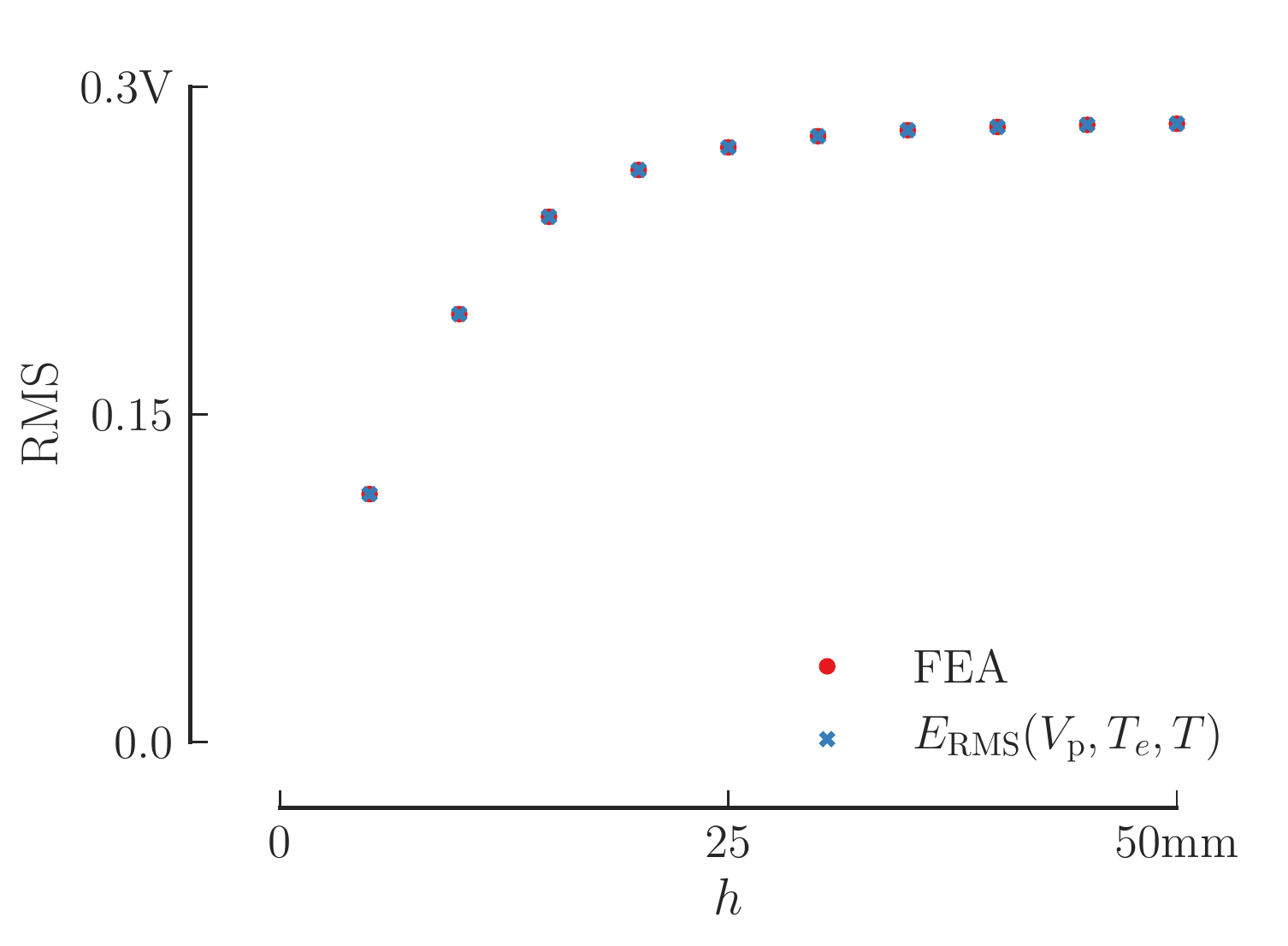}
  \caption{Comparison of \cref{eq:RMS_single_sine} to the RMS of open-circuit
    EMF simulated using FEA for discrete values of
    $h$. \label{fig:ErmsMaxwellComparison} }
\end{figure}

\Cref{eq:RMS_single_sine} can be used to approximate the open-circuit RMS EMF
obtained by FEA, shown in \cref{fig:oc_emf}. From
\cref{fig:ErmsMaxwellComparison}, it is clear that this approximation is
accurate.

Note that \cref{eq:RMS_single_sine,eq:oc_rms} both describe the same open
circuit RMS EMF, but as functions of different parameters. Since the parameters
$V_{\text{p}}$ and $T_e$ are dependent on $h$ and if the RMS calculations
leading to $e_{\text{RMS}}(h)$ by way of~\cref{eq:oc_rms} are performed over the
same length of time $T$ used in \cref{eq:RMS_single_sine}, then
\cref{eq:RMS_single_sine,eq:oc_rms} must produce the same RMS value. Hence, for
$T=T_1$ and $h=h_1$,

\begin{equation}
  e_{\text{RMS}}(h=h_1)|_{T=T_1} = e_{\text{RMS}}(V_{\text{p}}, T_e, T=T_1)|_{h=h_1}.
  \label{eq:1c1mRMSequal}
\end{equation}

\subsubsection{Multiple coil, single magnet configuration}\label{sec:mcsm}
We now extend the model for a single-coil single-magnet microgenerator to the
more general case of $c$ identical coils, with identical heights $h$, and a
single magnet $(m=1)$. As the magnet passes through each of the coils, an EMF of
the form shown in \cref{fig:oc_emf} will be induced. This basic waveform will
henceforth be referred to as the \textit{basic EMF pulse}. It has been shown to
be well approximated by two half-sine periods.

In addition, we can observe that, for a single-axis microgenerator, as an
individual magnet passes through a coil, it induces a basic EMF pulse in that
coil. Thus, the number of such pulses $ = c \cdot m$, where $c$ is the number of
coils and $m$ is the number of magnets present in the system.

Since our goal is to maximize the energy we can harvest, destructive
superposition among these sequential pulses should be avoided. For this to
occur, two requirements must be met. First, the polarity of the pulses must
match. This can be achieved by ensuring the correct sequence of coil polarities.
Second, adjacent coils must be spaced in such a way as to optimally superimpose
the lobes of successive EMF pulses. This can be achieved by suitable choice of
the length and spacing of the coils.

\begin{figure}[ht]
  \centering \includegraphics[width=0.6\textwidth]{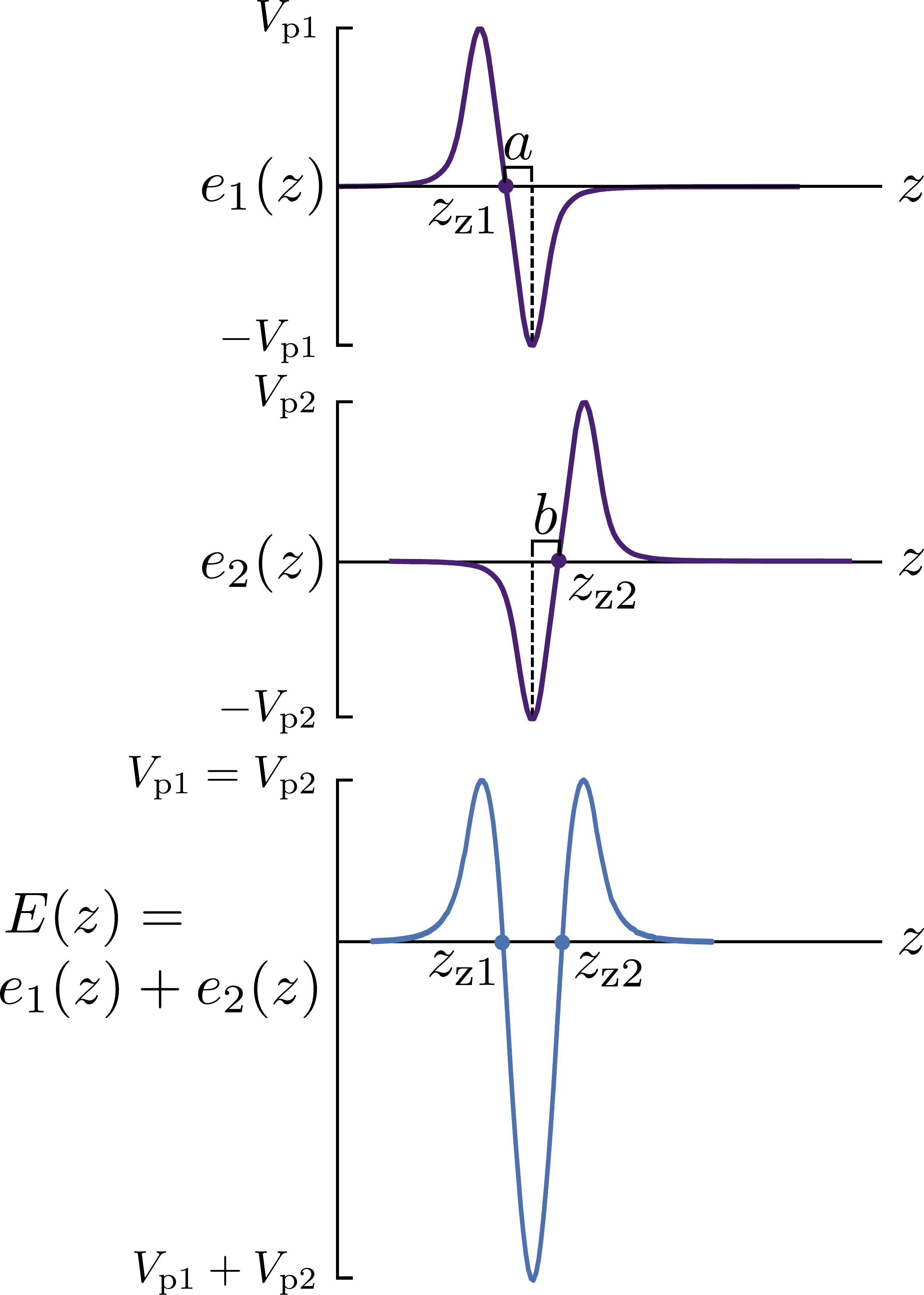}
  \caption{Two individual pulses, $e_1(z)$ with peak $V_{\text{p1}}$ and
    $e_2(z)$ with peak $V_{\text{p2}}$, produced by a device with $c=2$ coils
    and $m=1$ magnet (or alternatively $c=1$, $m=2$ as discussed in \cref{sec:mcsm}) and the optimal superposition of the two
    pulses , resulting in waveform $E(z)$. \label{fig:waveforms_indi} }
\end{figure}

Consider the EMF waveform $E(z)$ in \cref{fig:waveforms_indi}, which is the
result of a favourable superposition of two successive individual basic EMF
pulses $e_1(z)$ and $e_2(z)$ resulting from the single magnet passing through
two successive coils, and where $z$ is the vertical displacement of the magnet.
The polarities of the pulses have been matched to ensure that the trailing
negative excursion of the first pulse is reinforced by the leading negative
extrusion of the second. Since the two coils are identical $V_{\text{p1}} =
V_{\text{p2}}$. Furthermore, as can also be seen from \cref{fig:waveforms_indi},
the pulse peaks $V_{\text{p1}}$ and $V_{\text{p2}}$ coincide when the
displacement between the zero-crossing (indicated by $z_{\text{z1}}$ and $z_{\text{z2}}$) of the pulses is

\begin{equation}
  z_{\text{z2}} - z_{\text{z1}} = a+b,
  \label{eq:dist_max_power}
\end{equation}
where $a$ is the $z$-distance between $z_{\text{z1}}$ and the trailing peak of
$e_1(z)$, and $b$ is the $z$-distance between $z_{\text{z2}}$ and the leading
peak of $e_2(z)$.

\begin{figure}[ht]
  \centering \includegraphics[width=0.5\textwidth]{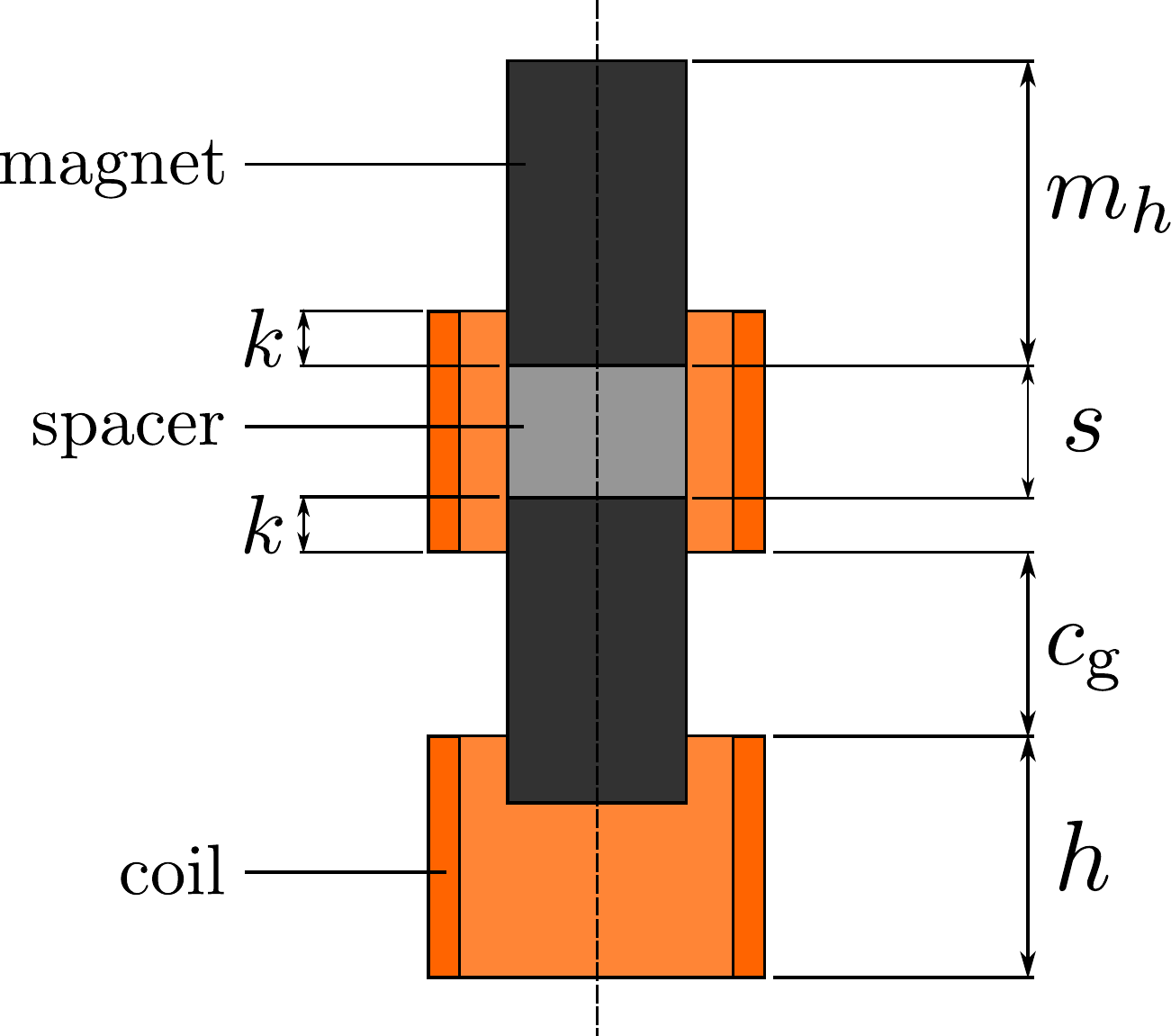}
  \caption{Cross section showing a position of the magnet assembly when it
    induces one of the EMF waveform peaks. This occurs when the leading or
    trailing edges of the magnets is a distance $k$ from the edge of the coils.
    The lengths $m_{\text{h}}, s, h$ and $c_{\text{g}}$ denote the magnet
    height, magnet spacer height, coil height and coil gap,
    respectively. \label{fig:multi_coil}}
\end{figure}

The use of the zero-crossings as a frame of reference is deliberate, as their
positions are independent of the coil properties and the magnetic field. This is
not true for the position of the peaks, for example. The zero-crossings occur
when the rate of change of flux $\frac{d\phi}{dt}=0$, which occurs when the
centre of the magnet passes through the centre of the coil. Consider
\cref{fig:multi_coil}, where parameter $k$ indicates the position at which the
peaks of a pulse occur and correspond to the displacement between the leading and
trailing coil edges and the displacement between leading and trailing magnet edges.
This allows \cref{eq:dist_max_power} to be related to the dimensions shown in
\cref{fig:multi_coil}, as follows

\begin{equation}
  z_{\text{z2}} - z_{\text{z1}} = c_{\text{g}} + h,
  \label{eq:multi_coil_spacing}
\end{equation}
where $c_{\text{g}}$ is the gap between the coils in mm and is given by
\begin{equation}
  c_{\text{g}} = m_{\text{h}} - 2k.
  \label{eq:coil_gap}
\end{equation}

Since the height $h$ of all coils are identical,
\cref{eq:multi_coil_spacing,eq:coil_gap} will hold for any number of coils when
there is a single moving magnet ($m=1$).

Finally, the total internal resistance of the microgenerator $R_{\text{mcrg}}$ is calculated by multiplying the
resistance of a single coil $R_{\text{coil}}$, given by~\cref{eq:r_coil_param},
by the number of coils $c$:

\begin{equation}
  R_{\text{mcrg}} = c \beta \gamma h. \label{eq:R_mcrg}
  \end{equation}

  \subsubsection{Single coil, multiple magnet configuration} \label{sec:scmm}
  The previous section extended the model for a single-coil single-magnet
  microgenerator to one with $c$ identical coils and one magnet ($m=1$). The
  case of a single coil ($c=1$) and $m$ identical magnets with identical heights
  $m_{\text{h}}$ can be treated in an analogous way. In this case, the RMS of
  $E(z)$ is maximized by separating adjacent magnets in the magnet assembly with
  an iron spacer, and by alternating the polarity of these magnets. As in
  \cref{sec:mcsm}, this leads to opposite polarity for successive pulses. This
  leads to the requirement:
   
\begin{equation}
  z_{\text{z2}} - z_{\text{z1}} = s+m_{\text{h}},
  \label{eq:multi_mag_spacing}
\end{equation}
where $s$ is the spacing between the magnets and $m_{\text{h}}$ is the magnet
height. The value of $s$ is given by,
\begin{equation}
  s = h - 2k,
  \label{eq:mag_gap}
\end{equation}
where once again $k$ can be determined as described on \cref{sec:mcsm}.

Since the height of the magnets $m_{\text{h}}$ in the magnet assembly are
identical, \cref{eq:multi_mag_spacing,eq:mag_gap} will hold for any number of
magnets.

\subsubsection{Multiple coil, multiple magnet configuration}
To obtain a model for a microgenerator with $c$ coils and $m$ magnets we
simultaneously enforce \cref{eq:multi_mag_spacing,eq:multi_coil_spacing} to obtain:

\begin{equation}
  s+m_{\text{h}} = c_{\text{g}}+h.
  \label{eq:coil_mag_gap_combined}
\end{equation}

Now, \cref{eq:coil_mag_gap_combined,eq:coil_gap,eq:mag_gap} can be used to
determine the spacing of the coils and magnets that optimizes the RMS
voltage. 

\subsubsection{Generalized power delivery by multiple coil, multiple magnet
  designs} \label{sec:general_power} We now consider the peak amplitudes of
$E(z)$, the total open-circuit output voltage produced by the linear kinetic
energy harvester as an assembly with $m$ magnets that traverses $c$ coils. For
ease of analysis, we will normalize these peak voltages by the peak amplitude
$V_{\text{p}}$ of the basic pulse, as shown in~\cref{fig:sin_within_global}. The
magnitude of these normalized voltages will be denoted by $P(n,c,m)$, where $c$
refers to the number of coils, $m$ the number of magnets and $n$ refers to each
peak's index within the peak sequence. Since all magnets have the same
geometry and are evenly spaced, and all coils have the same geometry and are
evenly spaced, inspection of~\cref{fig:waveforms_indi} allows us to state for
the general case:

\begin{align} 
  \intertext{For $m > c$:}
  P(n,c,m)  &= \left\{
              \begin{array}{ll} \label{eq:m>c}
                P(-n+\lambda, c, m)&: N_i \leq n < 0 \\
                2c &: 0 \leq n \leq \lambda\\
                2(m-n)-1 &: \lambda < n \leq N_f \\
                0 &: \text{otherwise,}
              \end{array}
                    \right.
                    \intertext{where} 
                    \lambda &= m-c-1 \nonumber \\
  N_f &= m-1 \nonumber\\
  N_i &= -c. \nonumber
\end{align}
\begin{align}
  \intertext{For $m = c$:}
  P(n, c,m) &= \left\{
              \begin{array}{ll} \label{eq:m=c}
                P(-n+1, c, m) &: N_i \leq n < 0\\
                2c-1 &: n = 0 \\
                2(c-n)+1 &: 0 < n \leq N_f \\
                0 &: \text{otherwise,}
              \end{array}
                    \right.
                    \intertext{where}
                    N_f &= c = m \nonumber\\
  N_i &= c-1 = m-1. \nonumber
\end{align}
\begin{align}
  \intertext{For $m < c$:}
  P(n, c,m) &= \left\{
              \begin{array}{ll} \label{eq:m<c}
                P(-n+\lambda, c, m) &: N_i \leq n < 0\\          
                2m &: 0 \leq n \leq \lambda\\
                2(c-n)-1 &: \lambda < n \leq N_f \\
                0 &: \text{otherwise,}
              \end{array}
                    \right.
                    \intertext{where}
                    \lambda &= c - m -1 \nonumber \\
  N_f &= c -1 \nonumber \\
  N_i &= - m. \nonumber
\end{align}

In \cref{eq:m>c,eq:m=c,eq:m<c}, $N_i$ and $N_f$ are the values of $n$ that indicate initial and final peak of
the peak sequence.

\begin{figure}[ht]
  \centering
  \includegraphics[width=\textwidth]{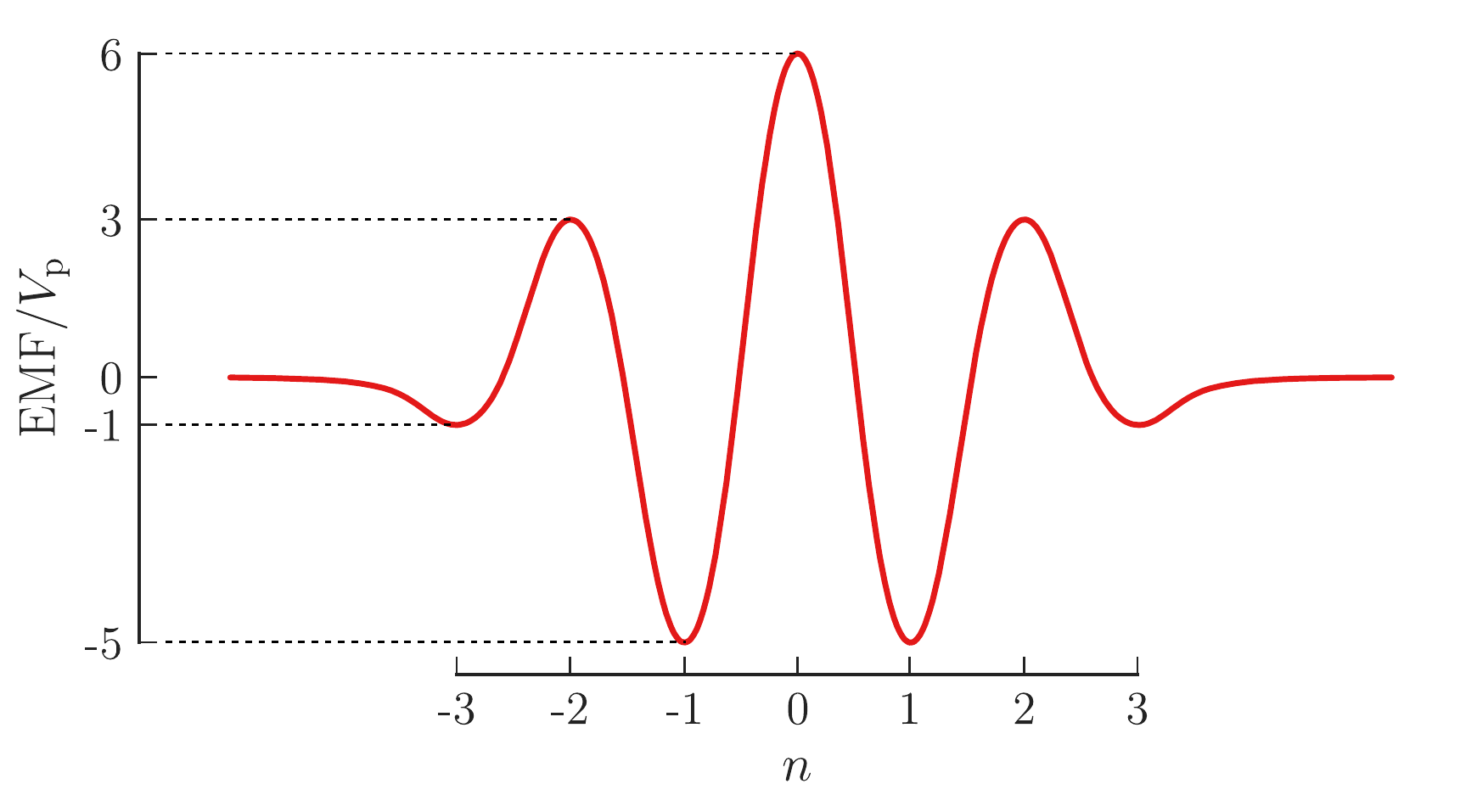}
  \caption{An example, showing the waveform produced by a microgenerator consisting of $c=3$
    coils and $m=3$ magnets, for a magnet assembly moving through the coils
    at a constant velocity. If we use \cref{eq:m<c}, as for this example $m<c$,
    we correctly predict the relative magnitude of peaks in the waveform to be
    $P(n,4,3) = \{1,3,5,6,5,3,1\}$ for each peak position $n \in \{-3,-2,-1,0,1,2,3\}$.\label{fig:waveform_c4m3} }
\end{figure}

A specific example , for a microgenerator consisting of $c=4$ coils and $m=3$
magnets is shown in \cref{fig:waveform_c4m3}. \Cref{eq:m<c,eq:m=c,eq:m>c} have been
used to calculate the magnitudes of the peaks produced by a microgenerator.

The true peak voltage is given by:

\begin{equation}
  \hat{P}(n,c,m) = |V_{\text{p}}| \cdot P(n,c,m).
  \label{eq:peaks_simple}
\end{equation}

This allows a comparison between kinetic microgenerators with different magnet
and coil properties, and hence different values of $V_{\text{p}}$.

Next we extend \cref{eq:RMS_single_sine}, which models the open-circuit RMS
voltage as a function of coil height $h$ for a single coil, single magnet
configuration, to the general case of $c$ coils and $m$ magnets. Since
$E(t)$ consists of a series of half-sine wave lobes, with each lobe's peak given
by \cref{eq:peaks_simple}, the open-circuit RMS voltage when $c$ coils and $m$
magnets are considered can be obtained by applying~\cref{eq:RMS_single_sine}: 
\begin{equation}
  E_{\text{RMS}}(c,m) = \sqrt{\frac{T_{e}}{4T} \sum_{n=N_i}^{N_f}\hat{P}(n, c,m)^2},
  \label{eq:E_RMS_general}
\end{equation}
where $T_{\text{e}}$ is the period of the basic pulse.

\Cref{eq:E_RMS_general} provides an approximation of the final open-circuit RMS
voltage delivered by the energy harvester in terms of design parameters $c$ and
$m$. However, since $T_e$ and $\hat{P}(n,c,m)$ depend on $h$, this result can be
developed further. We begin by defining the ratio of RMS open-circuit voltages
of microgenerators with different values of $c$ and $m$, but identical $h$ using
\cref{eq:E_RMS_general}:

\begin{align}
  \eta(c_1, m_1, c_2, m_2) &= \frac{E_{\text{RMS}}(c_2,m_2)|_h}{E_{\text{RMS}}(c_1,m_1)|_h} \nonumber \\
                           &= \sqrt{\frac{\sum_{N_{i2}}^{N_{f2}} P(n, c_2, m_2)^2}{\sum_{N_{i1}}^{N_{f1}} P(n, c_1, m_1)^2}}.
                             \label{eq:ratio}
\end{align}

Using
\cref{eq:oc_rms,eq:RMS_single_sine,eq:1c1mRMSequal,eq:ratio,eq:E_RMS_general}
and noting, from \cref{eq:m=c}, that $\sum_{N_{i1}}^{N_{f1}}P(n, 1,1)^2 = 1^2 +
1^2 = 2$ and by substituting into \cref{eq:power_avg}, an expression for the
average power delivered to the load can be determined:
\begin{equation}
  \bar{P}_{\text{load}}(h, c, m, R_{\text{load}}) = \frac{A^2R_{\text{load}} \cdot (1-\exp(-\alpha h))^2 \sum_{N_i}^{N_f}P(n,c,m)^2 }{2(R_{\text{load}}+ R_{\text{mcrg}})^2}.
  \label{eq:apg_pre}
\end{equation}

By substituting \cref{eq:R_mcrg}, the final expression is obtained as a function
of all design parameters.

\begin{equation}
  \bar{P}_{\text{load}}(h, c, m, R_{\text{load}}) = \frac{A^2R_{\text{load}} \cdot (1-\exp(-\alpha h))^2 \sum_{N_i}^{N_f}P(n,c,m)^2 }{2(R_{\text{load}}+ c \beta \gamma h)^2}.
  \label{eq:apg}
\end{equation}

We note again here that \Cref{eq:apg} is based on an idealized motion in which
the magnet assembly moves through the coils at a constant velocity. Hence this
result does not necessarily accurately reflect the power that will be produced
by the device in practice. It is, however, expected to indicate which
microgenerator among competing designs will produce maximum power during
practical operation. 

We base this assumption on the observation that the magnet assemblies of
different microgenerator designs move through the coils with highly similar
velocity profiles. Thus, while the result of the predicted power given by
\cref{eq:apg} may change for different acceleration impulses and velocities, the
ranking of the microgenerator devices based on their expected power output will
not. As a result, optimizing for idealized motion can be used as a substitute
for optimizing for non-idealized motion in our application.


%% file: designApplication.tex
\section{Design application}\label{sec:exp_eval}
We now apply the methods presented in the previous section to the design of a
microgenerator that harvests the maximum amount of energy from the walking
motion of a human test-subject. 

\subsection{Physical constraints}\label{sec:constraints}
Due to a strong emphasis on size limitations in our intended eventual
application, axially-magnetized cylindrical N35 grade neodymium iron boron
(NdFeB) magnets were selected. The magnets have a radius of 5mm and a height of
5mm, and were the strongest readily available at the time. In order to increase the magnetic
flux density, two magnets are placed together with poles aligned, resulting in an
effective height of 10mm.

Initial experimentation found than an inner tube radius that was 0.5mm greater
than the magnet radius ensured unhindered motion of the magnet assembly. The
microgenerator tube body was 3D printed using PLA with a thickness of 1mm.

AWG36 gauge copper wire was selected for the coil, giving a wire radius
of $r_c = 0.0635$mm and a resistance per unit length of $R_{\text{gauge}}=1361
\times 10^{-6}~ \Omega/\text{mm}$. The small wire gauge allowed a large winding
density close to the magnet assembly while minimizing the coil diameter. The
resistance per unit length of the wire was also taken into consideration, as its higher
value allows flexibility in load-matching by varying $h$, and hence
$R_{\text{mcrg}}$, as per \cref{eq:R_mcrg}.

The horizontal coil thickness $c_{\text{th}}$ plays a significant role in the
open-circuit EMF that is induced. While this does present another possible
avenue for optimization, the effect of $c_{\text{th}}$ is currently not
explicitly modelled as a parameter. Instead, by selecting a $c_{\text{th}}$
value, its effects will be modelled implicitly in~\cref{eq:oc_rms}. We
determined a suitable value of $c_{\text{th}}$ by FEA, whereby the value of
$c_{\text{th}}$ is large enough to provide a turn density $\gamma$ and
resistance per turn $\beta$ that allows for increased EMF and load
impedance-matching capability by varying the coil height $h$, without resulting
in an excessively large internal coil resistance
$R_{\text{coil}}$ and severe power losses in the microgenerator. This resulted
in a horizontal coil thickness of $c_{\text{th}}=0.725$mm, which serves as a
compromise between these two extremes. The bottom thickness of the tube and the
thickness of the tube lid was selected as $t_{\text{up}}=t_{\text{lo}}=2$mm.
  
\begin{figure}[ht]
  \centering \includegraphics[width=0.7\textwidth]{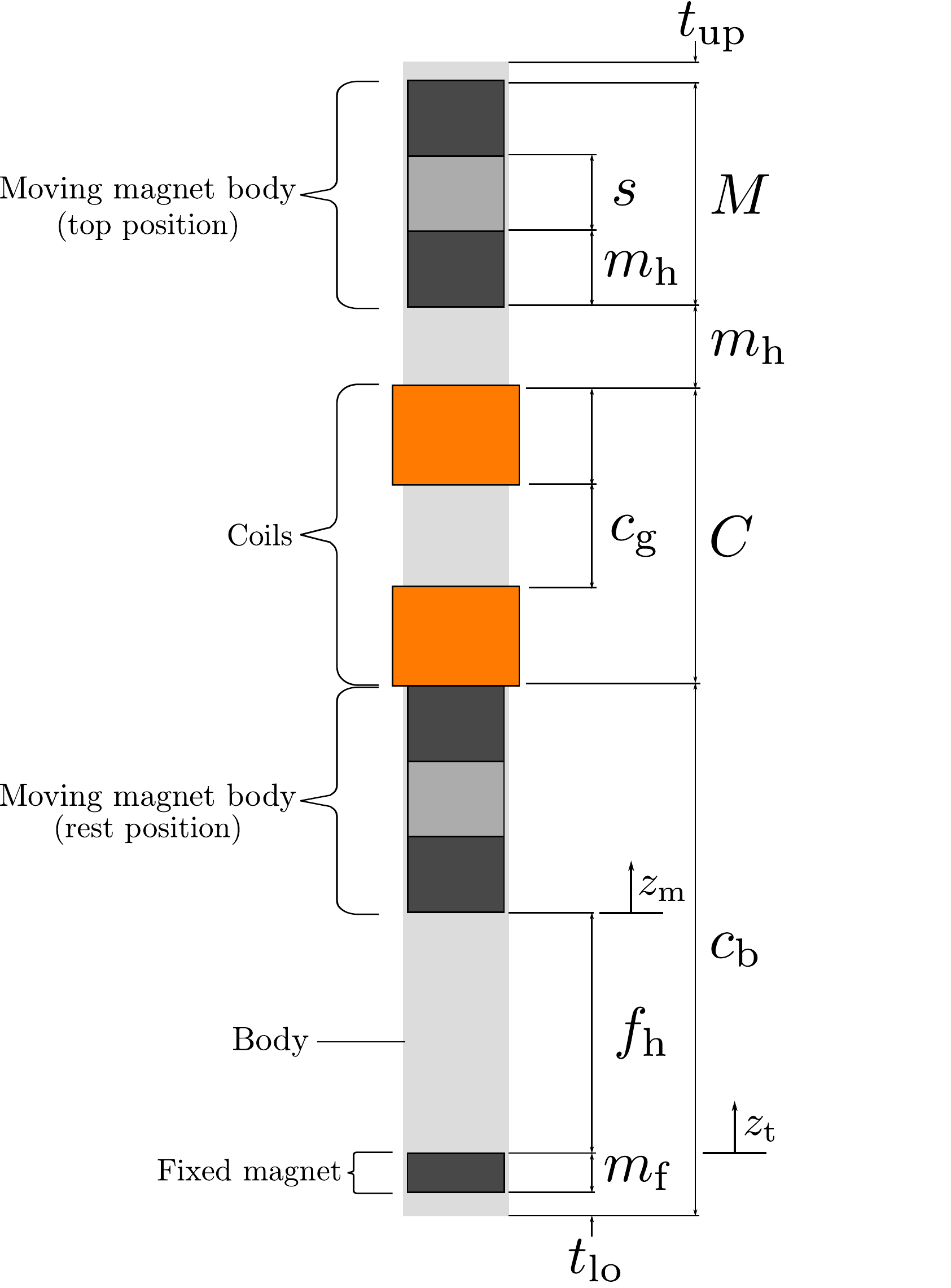}
  \caption{Vertical dimensions of kinetic microgenerator body, showing the
    required range of motion of the magnet assembly. \label{fig:total_height} }
\end{figure}

The maximum vertical space available for the microgenerator is constrained to
$L=125\si{mm}$. Using dimensions as defined in \cref{fig:total_height}, the
total height is given by
\begin{equation}
  L = 2M + C + m_{\text{h}} + m_{\text{f}} + f_{\text{h}} +t_{\text{up}} + t_{\text{lo}} \label{eq:L_total}
\end{equation}
where
\begin{align}
  M &= m \cdot m_{\text{h}} + (m-1)s \label{eq:M} \\
  C &= c \cdot h + (c-1)c_{\text{g}} \label{eq:C} \\
  f_{\text{h}} &= \delta^{-1}_{\text{mag}}(w_{\text{M}}) \label{eq:floating_height},
\end{align}
and $w_{\text{M}}$ is the weight of the magnet assembly. An expression for the
magnet assembly floating height $f_\text{h}$ is obtained by considering the
microgenerator at static equilibrium, where the magnetic spring force
$\delta_{\text{mag}}(z)$ is equal to the weight of the magnet assembly. As shown
in~\cref{fig:total_height}, the floating height is defined the distance between
the magnet assembly and the bottom magnet. Using \cref{eq:floating_height}, the
floating height $f_{\text{h}}$ can be expressed as:

\begin{equation}
  \delta_{\text{mag}}(f_{\text{h}}) = w_{\text{M}}, \label{eq:fh_pre}
\end{equation}

Thus the vertical height is constrained by
\begin{equation}
  2M + C + m_{\text{h}} + m_{\text{f}} + f_{\text{h}} +t_{\text{up}} + t_{\text{lo}} \leq 125\si{mm}.
  \label{eq:L_constraint}
\end{equation}

For the magnet assembly to move through all the coils, sufficient range of vertical
motion is required. This places an additional constraint on the minimum height
of the device. The minimum range of relative motion required is
\begin{equation}
  z_{\text{m}|\text{t,top}} - z_{\text{m}|\text{t,rest}}  \geq M + C + m_{\text{h}},
  \label{eq:motion_constraint}
\end{equation}
where $z_{\text{m} | \text{t}}=z_m - z_t$ as shown in \cref{fig:total_height}.

Experimentation indicated that the magnet assembly cannot be reliably
constructed for a spacer height less than $s<s_{\text{min}}=2.5\text{mm}$. This places an
implicit constraint on $h$, which we can derive from \cref{eq:mag_gap},

\begin{align}
  h \geq s_{\text{min}} + 2k \label{eq:min_coil_general}
\end{align}

\Cref{eq:L_total,eq:M,eq:C,eq:floating_height,eq:motion_constraint,eq:min_coil_general}
now provide a set of parameterized constraints that can be used to produce a set
of optimized microgenerator configurations that conform to the practical
limitations of the application.

\subsection{Parameter calculation and optimization}
\label{sec:calculation} We are now in a position to calculate the parameters of
the mechanical system model and the parameters of the electrical system model.
Using both sets of parameters, full optimization of the microgenerator is
performed. This results in a number of possible architectures for subsequent
comparison and assessment.

\subsubsection{Mechanical system model} \label{sec:phys_sys_mod} For the
mechanical system model, the magnetic spring force $\delta_{\text{mag}}(z)$ is
found by determining the force between two magnets at discrete points using FEA
and then fitting \cref{eq:coulombs_modified} to these values. For our choice of
magnets this results in the following relation for $\delta (z)$, where $z$ is in
metres.

\begin{equation}
  \delta(z) = \frac{\num{2.943e-3}}{4 \pi z^2 + \num{1.125e-4}}, 
\end{equation}

\begin{figure}
  \begin{subfigure}[b]{\textwidth}
    \captionsetup{justification=justified,singlelinecheck=false}
    \caption{\label{fig:footstepdata}}
    \includegraphics[width=\linewidth]{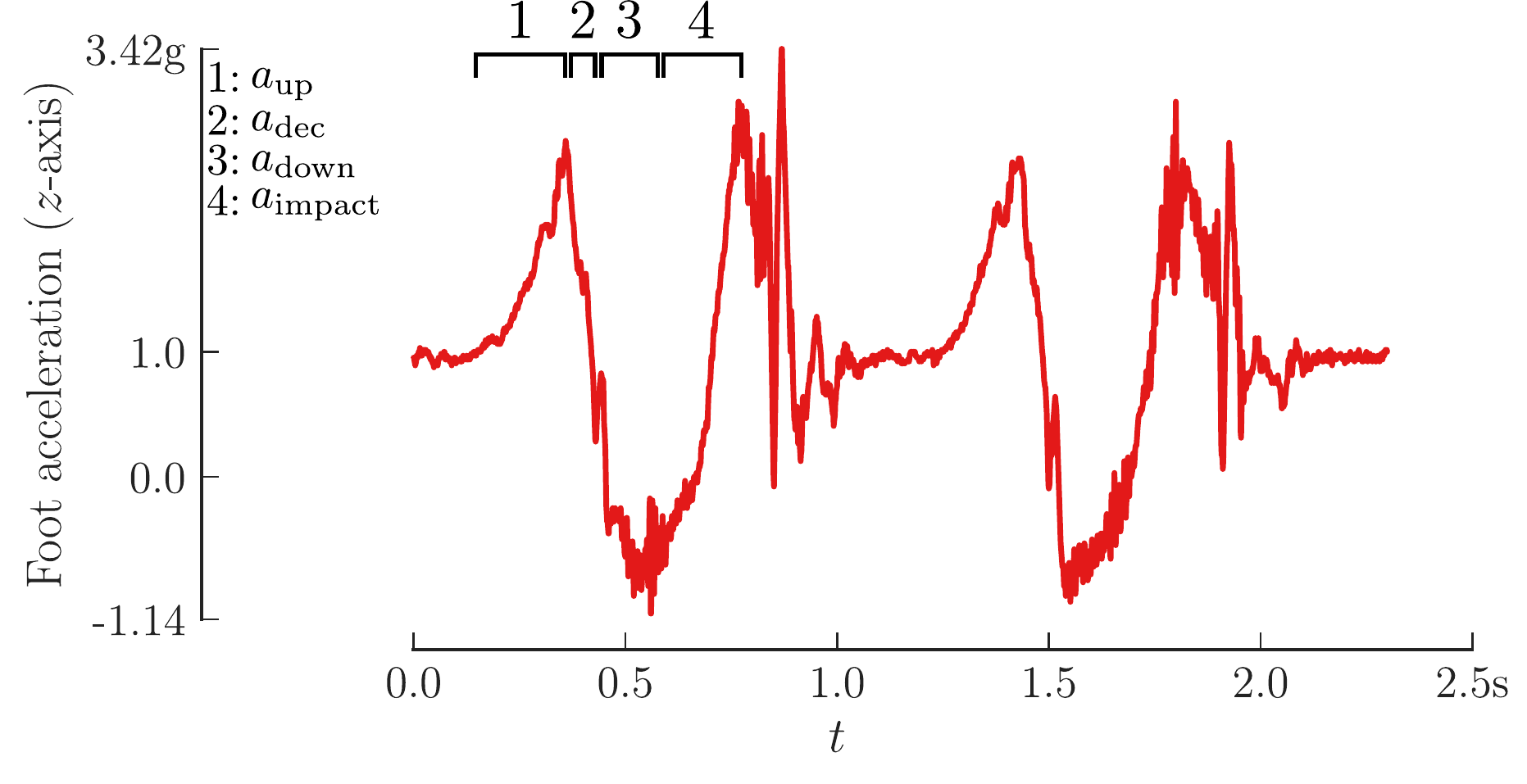}
  \end{subfigure}

  \begin{subfigure}[b]{\textwidth}
    \captionsetup{justification=justified,singlelinecheck=false}
    \caption{\label{fig:foot_gen}}
    \includegraphics[width=\linewidth]{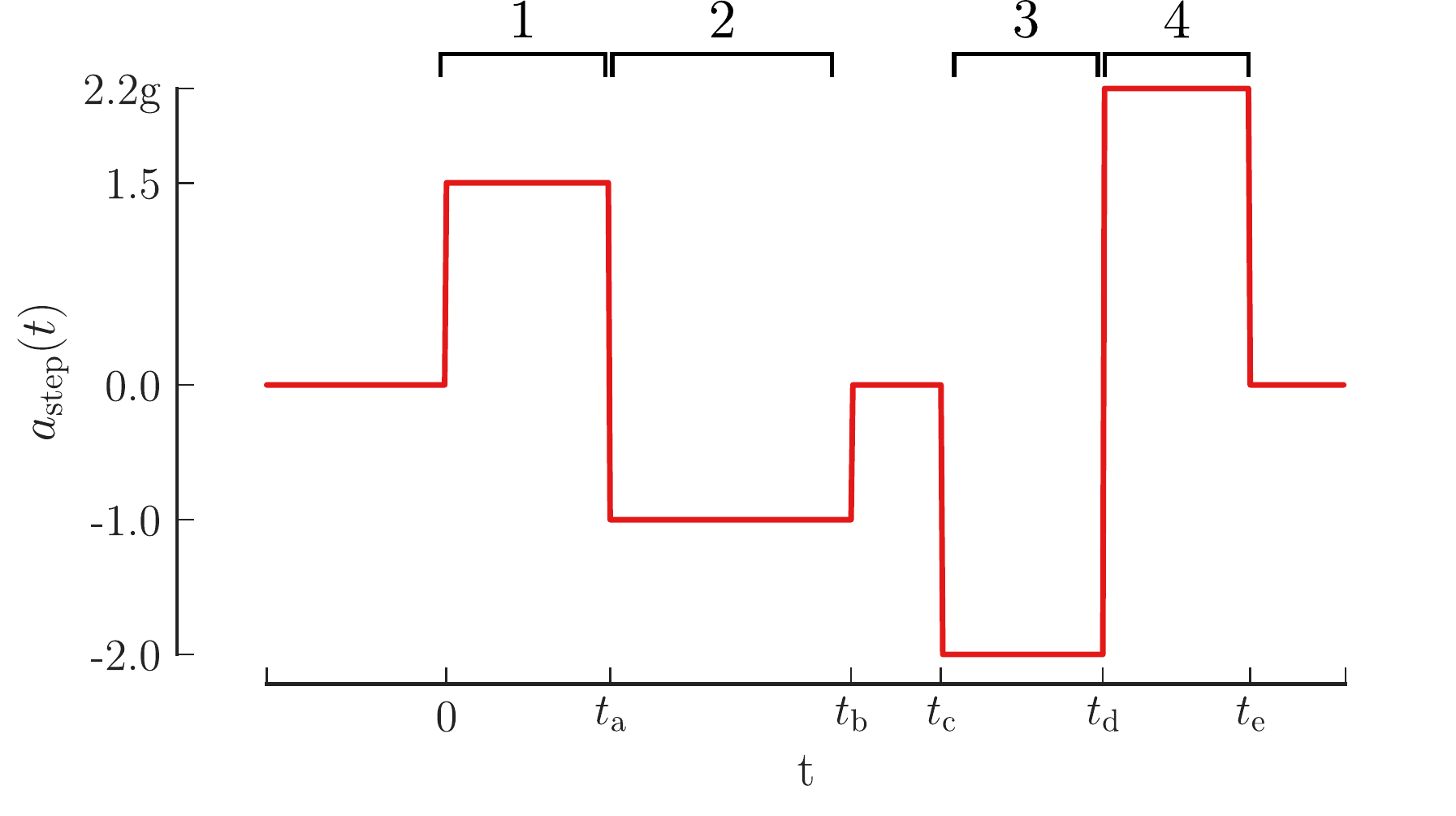}
\end{subfigure}

\caption{Comparison of accelerometer measurements from the leg of a human (in
  the $z$-direction) with artificial footstep described in \cref{sec:model}. (a)
  Typical accelerometer measurements from leg of human in the
    $z$-direction. The foot experiences upward accelerations of approximately
    2g, downard accelerations of approximately 2g and impact accelerations of
    2.2 to 2.42g, after removing the residual gravity measured by the
    accelerometer. The footstep stages, described in~\cref{fig:footstep_extra}, are
    indicated. (b) Artificial footstep used as input for physical system model, with
    corresponding time instances specified in \cref{tab:footstep_times}. This is
    a human-specific instance of the general footstep model proposed
    in~\cref{fig:footstepSymbols}.
}
\end{figure}

The footstep acceleration function $a_{\text{step}}(t)$ is determined from
measurements taken at the lower leg of a walking human, a sample of which is
shown in \cref{fig:footstepdata}. From video footage, the maximum vertical
displacement of the leg was determined to be $s_{\text{h}}=0.15\si{m}$ and the
upstroke-downstroke delay to be approximately $t_{\text{delay}}=0.05\si{s}$. The
values of the footstep accelerations, $a_{\text{step}}(t)$, are calculated using
\cref{eq:time_a,eq:time_b}. The resulting function is shown in
\cref{fig:foot_gen}.

\begin{table}[ht]
  \centering \sisetup{ table-number-alignment = right, table-figures-integer =
    1, table-figures-decimal = 3 }
  \begin{tabular}{@{}lS@{}}
    \toprule
    $t$ & {seconds} \\
    \midrule
    $t_{\text{start}}$ & 0 \\
    $t_{\text{a}}$ &  0.091 \\
    $t_{\text{b}}$ &  0.225 \\
    $t_{\text{c}}$ &  0.275 \\
    $t_{\text{d}}$ &  0.365 \\
    $t_{\text{e}}$ &  0.447 \\
    \bottomrule
  \end{tabular}
  \caption{Time intervals corresponding to selected artificial footstep footstep
    function, shown in \cref{fig:foot_gen}. \label{tab:footstep_times}}
\end{table}

Finally, the damping coefficient $b_{\text{damper}}$ is experimentally determined as
 
\begin{equation}
  b_{\text{damper}} = 150 \cdot S_{M},
  \label{eq:}
\end{equation}

where $S_{\text{M}}$ is the outer surface area of a cylinder, excluding end
caps, of the magnet assembly.
\begin{figure}[ht]
  \centering \includegraphics[width=0.8\textwidth]{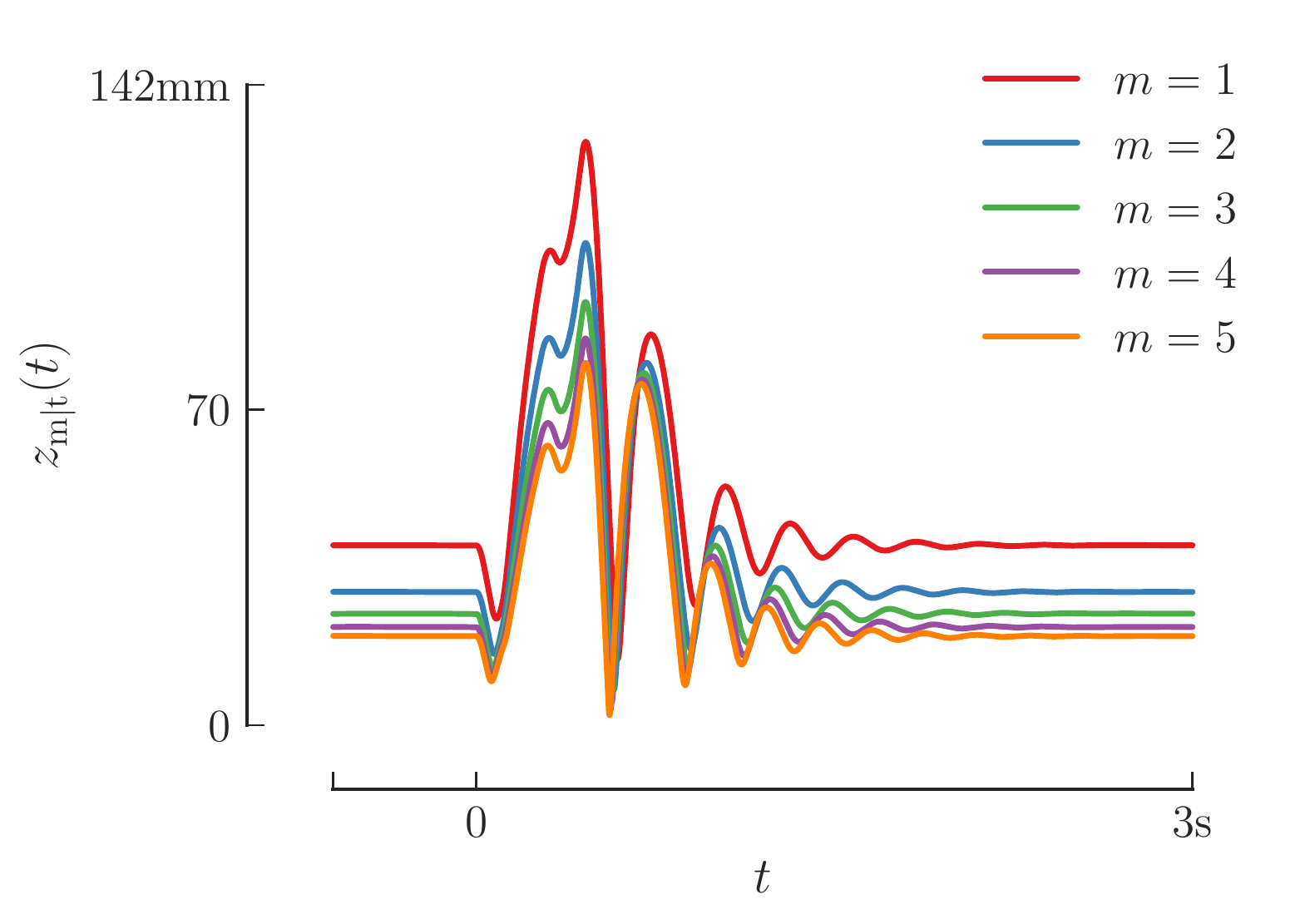}
  \caption{Numerical solution of \cref{eq:mech_system} for the relative
    displacement $z_{\mathrm{m|t}}(t)$ between the magnet assembly and the
    microgenerator tube with input shown in \cref{fig:foot_gen} for different
    number of magnets $m$. \label{fig:motion_solve} }
\end{figure}

By numerically solving for the parameters of our mechanical system model, given
by~\cref{eq:mech_system}, the motional constraint given by
\cref{eq:motion_constraint} can be found for different numbers of magnets $m$. This
constraint can then be imposed during architecture optimization.
\Cref{fig:motion_solve} shows the numerical solution of the mechanical system
model and indicates the predicted relative displacement between the magnet
assembly and the microgenerator tube for different values of $m$.

\subsubsection{Electrical system model} \label{sec:elec_sys_model2} The values
of $r_{\text{c}}$ and $R_{\text{gauge}}$ were selected in
\cref{sec:constraints}, allowing coil properties $\gamma$ and $\beta$ to be
calculated using \cref{eq:turn_density,eq:ohm_per_turn}, respectively. In our
specific case, this gives
\begin{align}
  \beta &\approx 0.059227 ~\Omega \text{/turn} \label{eq:beta_value}  \\
  \gamma &\approx 40.06 ~ \text{turns/mm}. \label{eq:gamma_value}
\end{align}

Recall that \cref{eq:turn_density,eq:ohm_per_turn} are two key coil properties
that allow for the coil resistance of the microgenerator $R_{\text{mcrg}}$ to be
estimated. As the coil height $h$ and the number of coils $c$ varies during the
optimization process, this too leads to variation in $R_{\text{mcrg}}$ during
optimization. Since the estimation of the load power is dependent on
$R_{\text{mcrg}}$, it must be known at each optimization step. This is made
possible by substituting \cref{eq:beta_value,eq:gamma_value} into \cref{eq:apg}
during the optimization process.

Next, FEA is used to simulate the EMF pulse of a single-coil, single-magnet model with
coil properties given by \cref{eq:beta_value,eq:gamma_value} for a range of
discrete values of $h$. The RMS of the EMF is calculated for each pulse that is
produced. \Cref{eq:oc_rms} is subsequently fit to this RMS data, giving an
accurate approximation of the open-circuit RMS EMF as a continuous function of
$h$,

\begin{equation}
  e_{\text{RMS}}(h) \approx 0.2860(1-\exp(-0.1111h)).
  \label{eq:}
\end{equation}

FEA is also used to determine the point at which the magnet assembly induces the
EMF peaks, denoted with dimension $k$ in~\cref{fig:multi_coil}. The value of $k$
was found to vary with the coil height $h$ for the range of coil heights
tested ($h \leq 50\text{mm}$) with the following relationship:

\begin{equation}
  k = h/4
  \label{eq:k_as_h}.
\end{equation}

The final electrical system parameter that must be determined is the electrical
load $R_{\text{load}}$. In our case this consists of a full-wave bridge
rectifier and a BQ25504 energy harvester from Texas Instruments. Since the
dynamic behaviour of this device's input resistance is not specified, it was
measured.

It is expected that the kinetic microgenerator will produce an average waveform
pulse current between $5\text{mA} \leq i_{\text{input}} \leq 10\text{mA}$. As a
result, the electrical load was selected as $R_{\text{load}}=40\Omega$.

\subsection{Architecture optimization} \label{sec:archi_opt}
An optimal architecture for the kinetic
microgenerator can be determined by maximizing the value of \cref{eq:apg} in
terms of the parameters $c, m$ and $h$, given the physical constraints.

Since the number of coils and magnets are integers $c, m \in
\mathbb{Z}$, a grid of integer values of $c$ and $m$ are used to
search for the optimum solution,
with each cell of the grid treated as an independent, restricted optimization
problem of \cref{eq:apg}, with parameters $c$ and $m$ fixed. In each case,
an optimal value of the parameter $h$ is determined using the sequential least
squares programming algorithm (SLSQP)\cite{Kraft1988}.

The seven best configurations determined by this procedure and their
corresponding parameters are listed in \cref{tab:opti_results}, alongside a
control baseline architecture that consists of a single coil of height $h=10$mm
and a single magnet. The control represents a baseline for what would be
considered a naive or initial experimental design for a linear kinetic energy
harvester, which consists of a single coil and a single magnet where certain
design verification is performed using FEA \cite{Bedekar2009}, and is typically
found in commercially available energy-harvesting shaker flashlights.

\begin{table}[ht]
  \centering \sisetup{ table-number-alignment = right, table-figures-integer =
    1, table-figures-decimal = 6}
  \begin{tabular}{@{}lS[table-figures-exponent=2]rrS[table-figures-integer=2,table-figures-decimal=2]r@{}}
    \toprule
    {Design} & {$\bar{P}_{\text{load}}$ (W)} & $c$ & $m$ & {$h$ (mm)} & {$L_{\text{min}}$ (mm)} \\
    \midrule
    1 &1.98899e-3 & 2 & 2 & 9.72 & 124.67 \\
    2 &1.37992e-3 & 1 & 3 & 5.98 & 125.0 \\
    3 &1.17282e-3 & 1 & 2 & 12.97 & 115.78 \\
    4 &0.62719e-3 & 3 & 1 & 8.34 & 114.74 \\
    5 &0.6146e-3 & 4 & 1 & 6.97 & 125 \\
    6 &0.596775e-3 & 2 & 1 & 9.87 & 102.86 \\
    7 &0.39094e-3 & 1 & 1 & 12.97 & 91.03 \\
    Control & 0.3756e-3 & 1 & 1 & 10.00 & 88.06
  \end{tabular}
  \caption{Relative power and design parameters for the set of most optimal
    kinetic microgenerator designs.\label{tab:opti_results}}
\end{table}
The best design uses $c = 2$ coils each with a height of $h = 9.72\text{mm}$ in
conjunction with a magnet assembly consisting of $m = 2$ magnets.

At this point it must be recalled that the power values given by \cref{eq:apg}
are based on an idealized motion, in which the magnet assembly moves through the
coils with constant velocity. While the power estimate does not necessarily
accurately indicate the power that will be produced by the device in practice,
it is expected to be indicative of the microgenerator design that will produce
maximum power during normal operation. Hence these power figures will only be
viewed as a relative measure of power for use in comparison with other potential
designs.

\begin{figure}[h]
  \centering
  \includegraphics[width=0.8\textwidth]{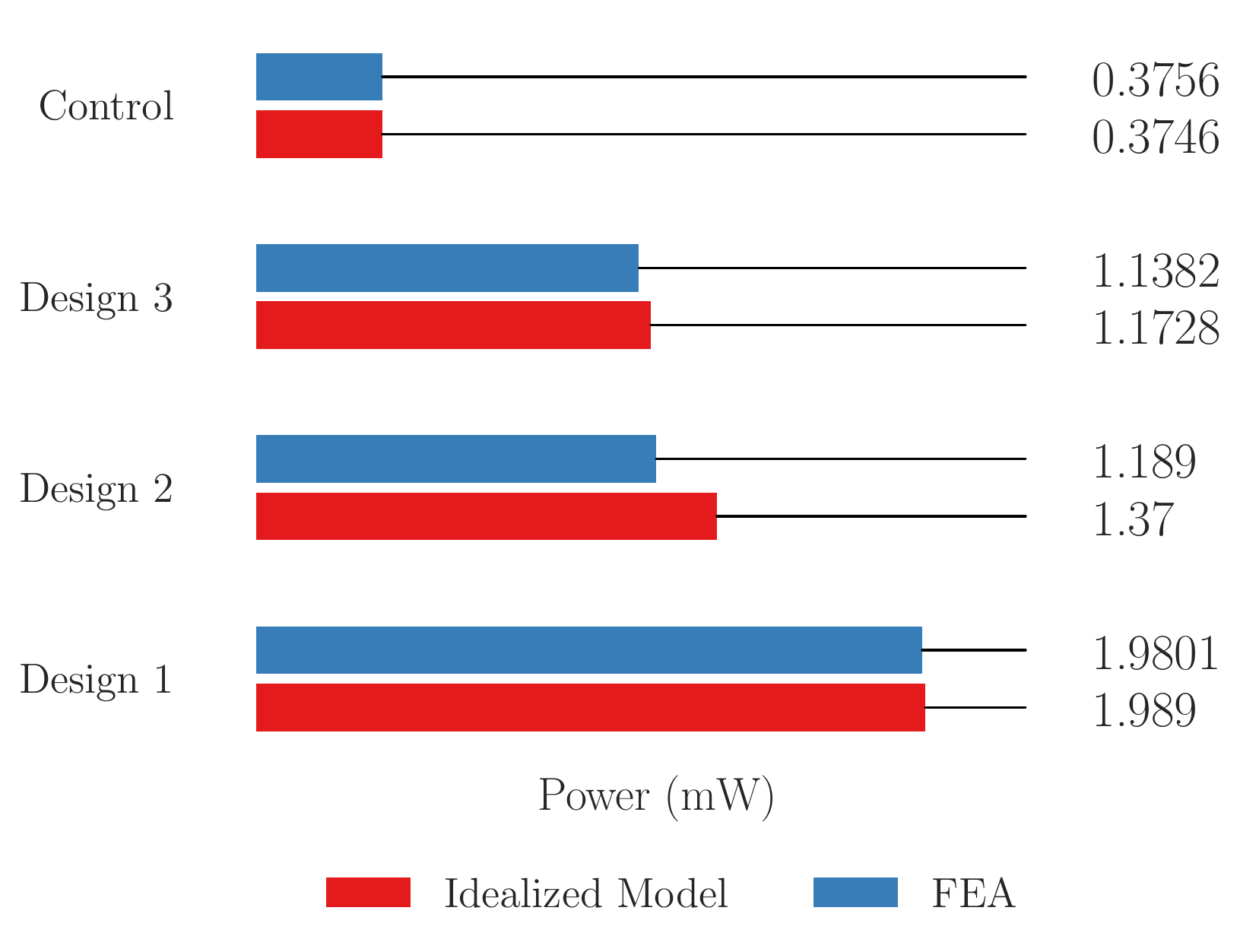}
  \caption{By comparing the results of the proposed model \cref{eq:apg} and
    the FEA results of $\bar{P}_{\text{load}}$, it is shown that the
    proposed model produces predictions with a high degree of
    accuracy. \label{fig:model_sim_comparison} }
\end{figure}

The microgenerator configurations in \cref{tab:opti_results} were assessed by
simulating the top three configurations and the control using FEA for idealized motion, and
calculating the power each design would deliver to $R_{\text{load}}$. The
results of this comparison are shown in \cref{fig:model_sim_comparison}.

From \cref{fig:model_sim_comparison} we see that the results produced by the idealized model described
by \cref{eq:apg} closely match those of the simulation. It
is interesting to note that the proposed model over-estimates the power by a
small degree for all configurations other than the control. This is due to the
assumption made by the proposed model that the leading and trailing edge of the individual
waveforms for the case of $c > 1$ and/or $m>1$, as shown in
\cref{fig:waveforms_indi}, do not overlap with the lobes of other waveforms.
This is an approximation, and the small degree of such overlap that does occur
in practice leads to a small degree of destructive interference. This is
expected to be more prominent for smaller values of $h$, because in this case
there is greater overlap between adjacent pulses. In addition, by treating the pulses as a series
of half-sine waveforms, the small EMF present prior to the first pulse and after
the last pulse is neglected. However,~\cref{fig:model_sim_comparison} indicates
that the effect of these factors is small across a diverse set of
microgenerator designs.


%% file: practicalTesting.tex
\section{Practical testing}
We now experimentally validate our previous assumption that a
microgenerator configuration produced by the proposed model and optimized for
idealized linear motion remains an optimal microgenerator in practice for more complex motion.

\subsection{Methodology}

\begin{figure}[h]
  \centering \includegraphics[width=\textwidth]{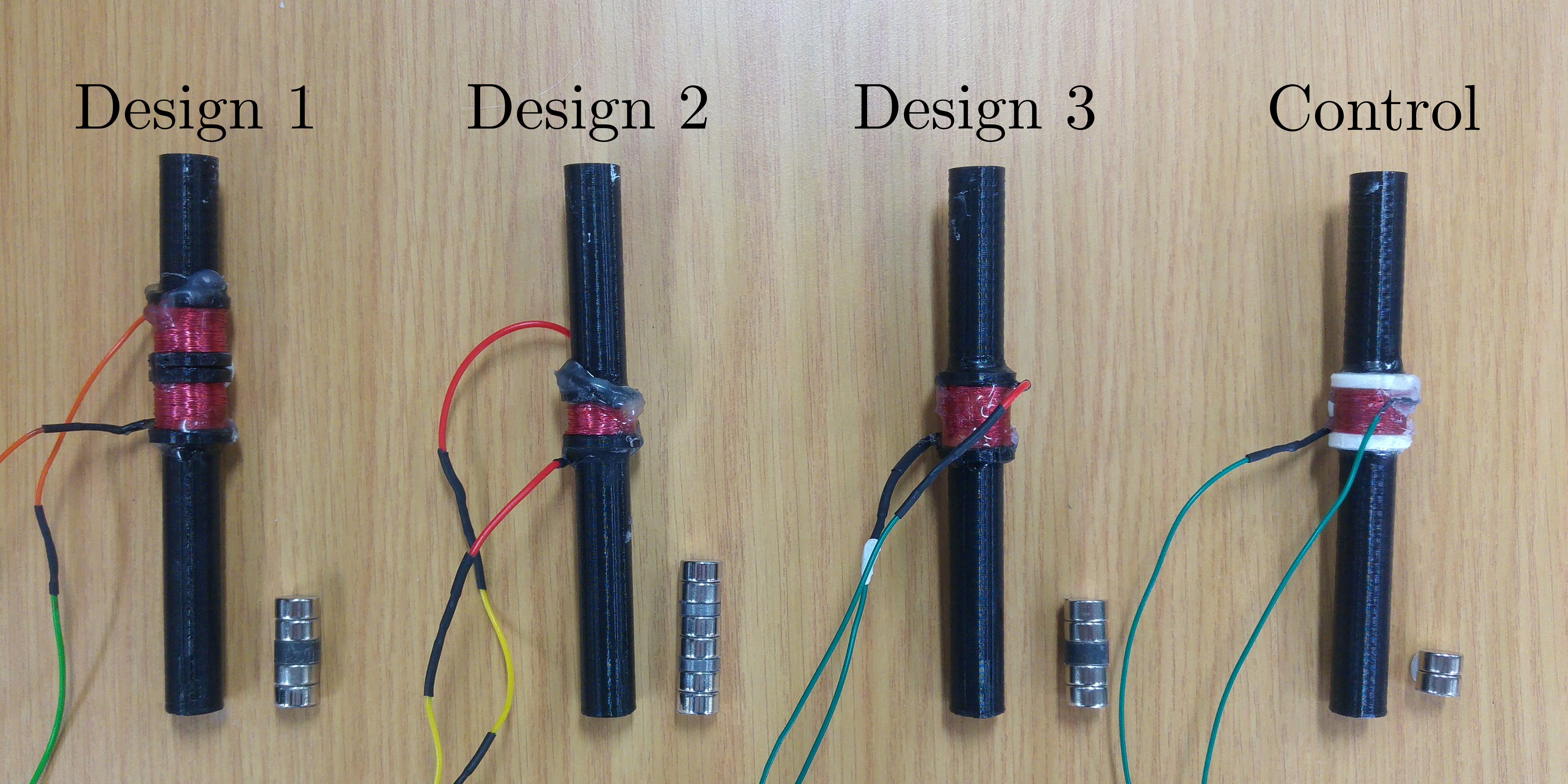}
  \caption{The top three and control microgenerator configurations and magnet
    assemblies were produced according to the specifications in
    \cref{sec:exp_eval}\label{fig:tubes_mag_assemblies}. }
\end{figure}

The microgenerator configurations shown in \Cref{fig:model_sim_comparison} were
built and assembled according to their specifications, with only a single
exception: the tube height for all configurations was set to $L=125\text{mm}$ to
ease the production process. Since the constraint on the tube height is
$L_{\text{min}} \leq L \leq 125\text{mm}$, with $L_{\text{min}}$ given for each
configuration in~\cref{tab:opti_results}, this has no effect on the test
outcome. The assembled microgenerators are shown alongside their magnet
assemblies in \Cref{fig:tubes_mag_assemblies}. The assembled generators each
have a total mass of 31.592g for Design 1, 32.673g for Design 2, 29.238g for
Design 3 and 17.248g for the Control.

\begin{figure}[h!]
  \centering
  \begin{subfigure}{0.3\textwidth}
    \centering \includegraphics[width=\textwidth]{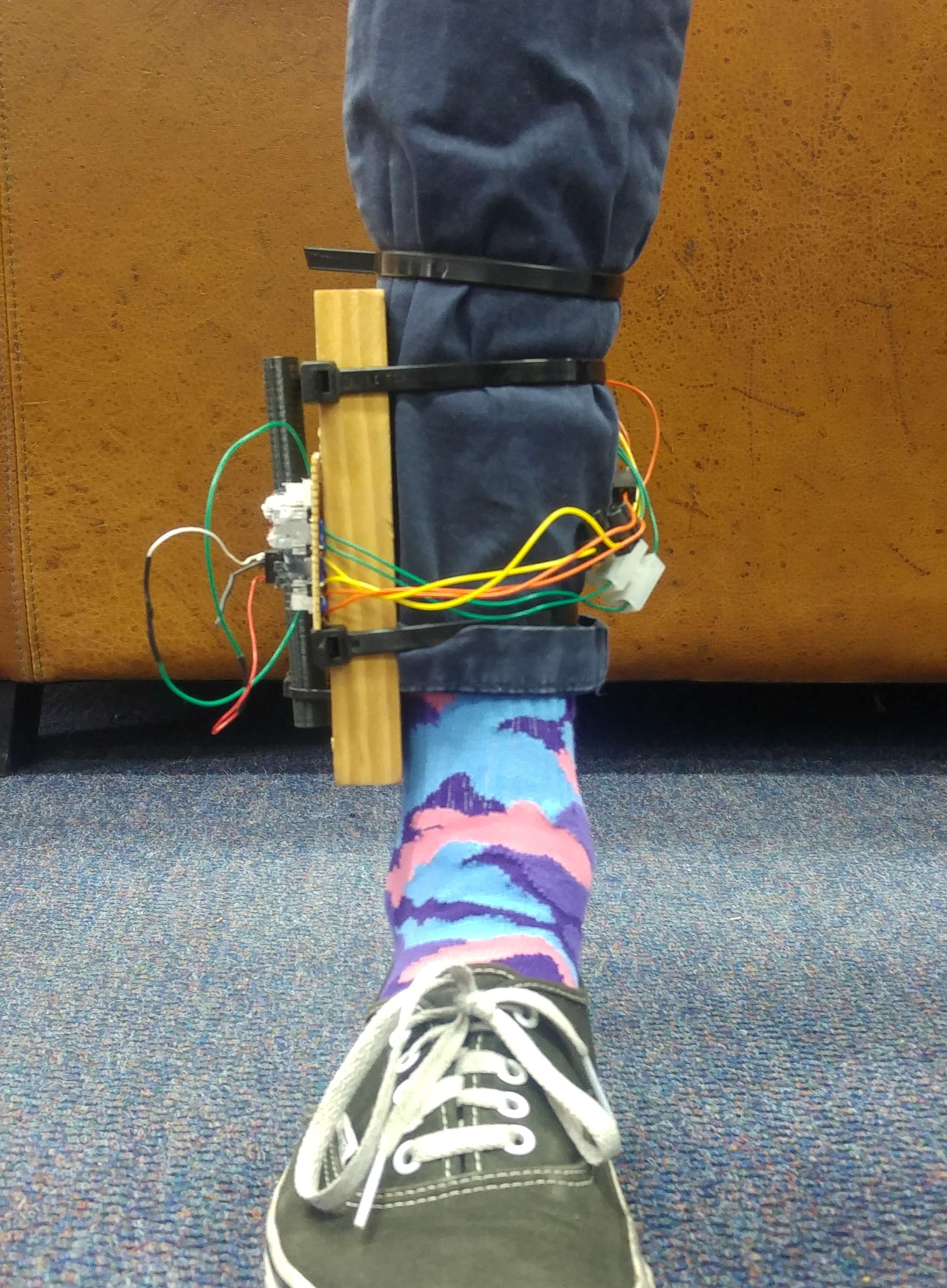}
    \caption{Front view.}
    \label{fig:test_front_view}
  \end{subfigure}
  \hfill
  \begin{subfigure}{0.3\textwidth}
    \centering \includegraphics[width=\textwidth]{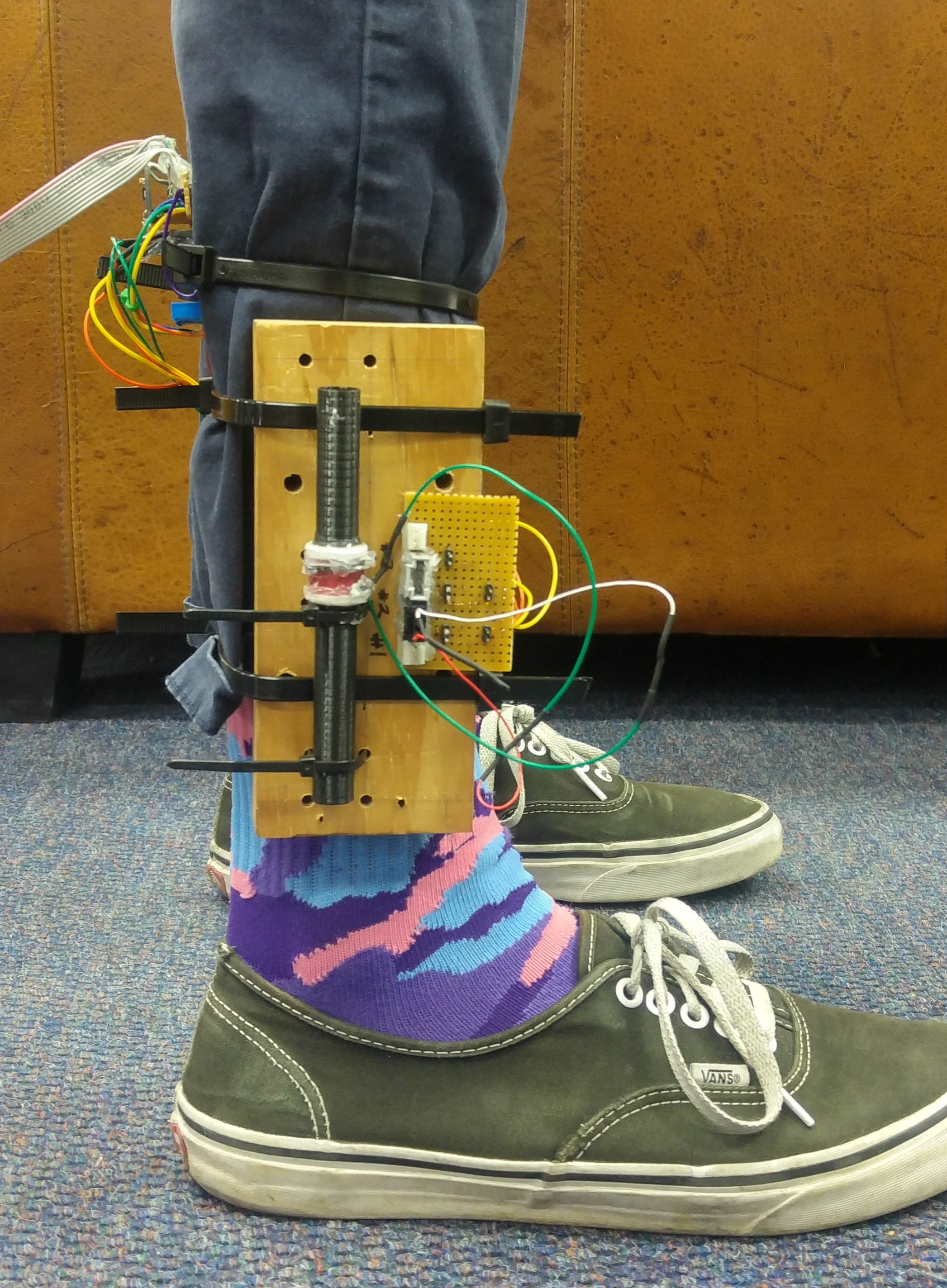}
    \caption{Side view.}
    \label{fig:test_side_view}
  \end{subfigure}
  \hfill
  \begin{subfigure}{0.3\textwidth}
    \centering \includegraphics[width=\textwidth]{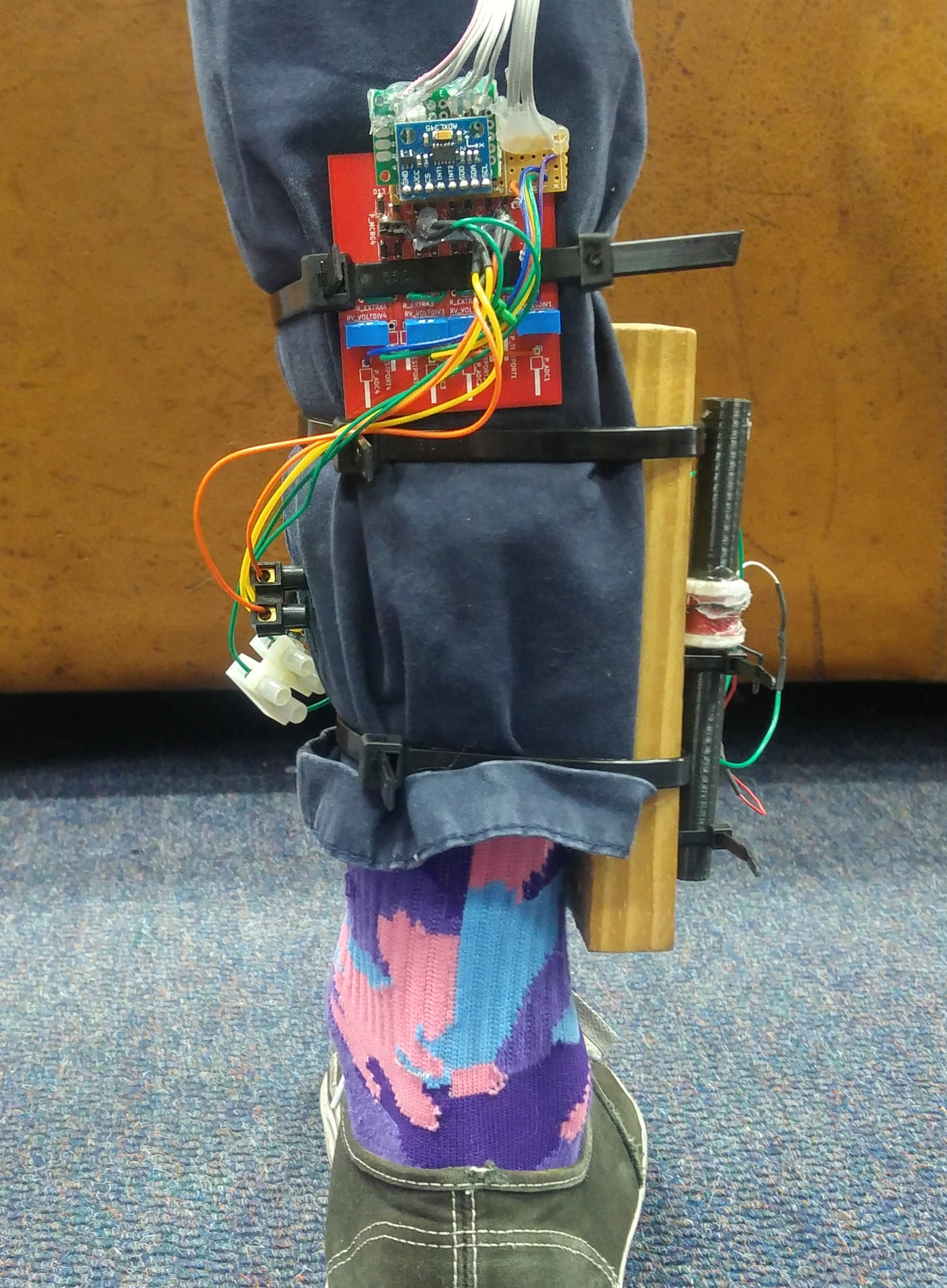}
    \caption{Rear view.}
    \label{fig:test_rear_view}
  \end{subfigure}
  \caption{For the human control test a mount that can hold a microgenerator
    device is attached the outer leg, with the voltage divider and accelerometer
    attached to the back of the leg.}
  \label{fig:test_human}
\end{figure}

\begin{figure}[h]
  \centering \includegraphics[width=\textwidth]{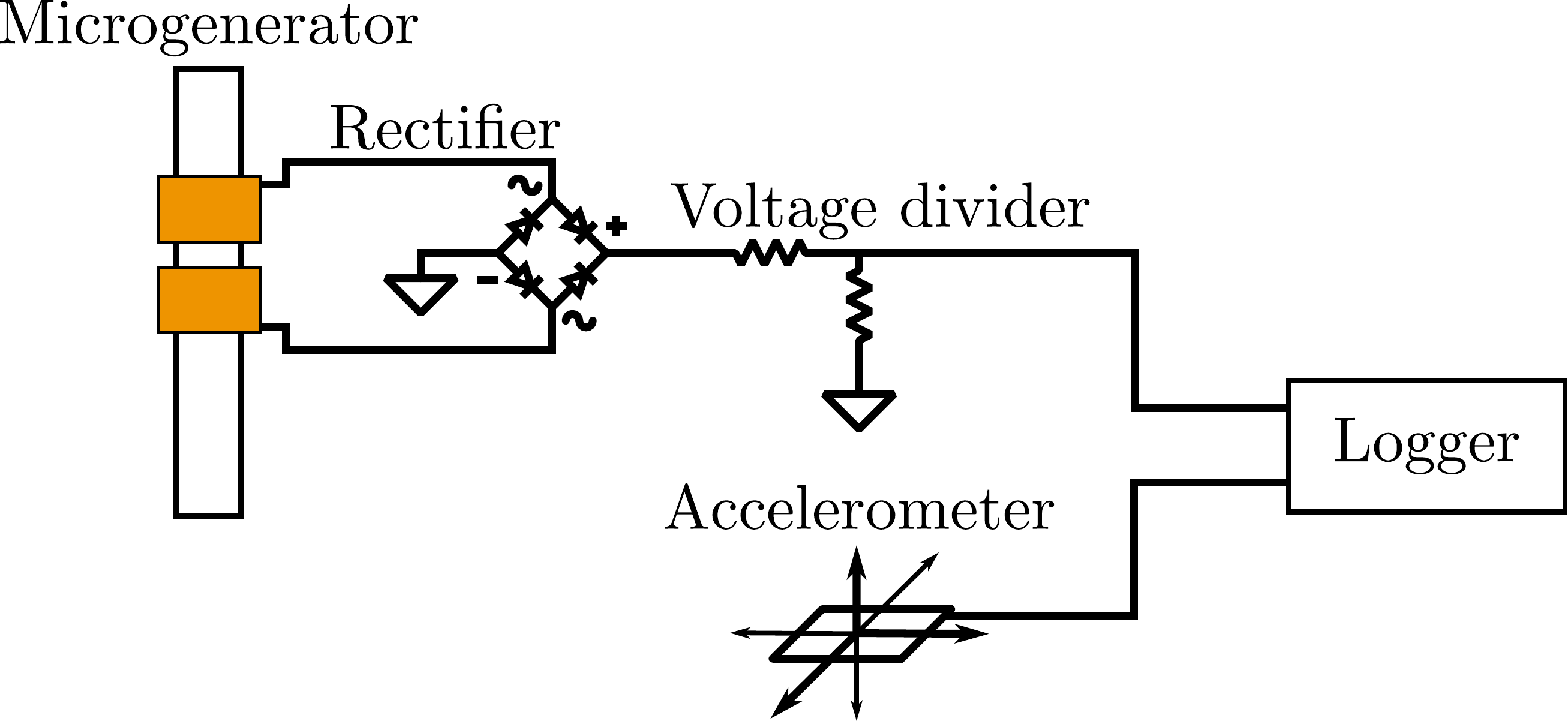}
  \caption{Simplified circuit diagram of the electronic circuit used for the
    test procedure. \label{fig:simplified_circuit}}
\end{figure}

A mount was fixed to the outer side of a human test subject's leg as shown in
\cref{fig:test_human}. The mount allows the microgenerator to be easily and
quickly swapped between tests, and ensures that it remains vertically aligned. A
full-wave bridge rectifier, voltage divider and accelerometer is attached the
leg and a data logger is held in-hand. A simplified circuit diagram is shown in
\Cref{fig:simplified_circuit}. The open-circuit EMF that is induced by the
microgenerator is rectified, scaled and logged synchronously with the
accelerometer output.

\begin{figure}[h]
  \centering \includegraphics[width=0.8\textwidth]{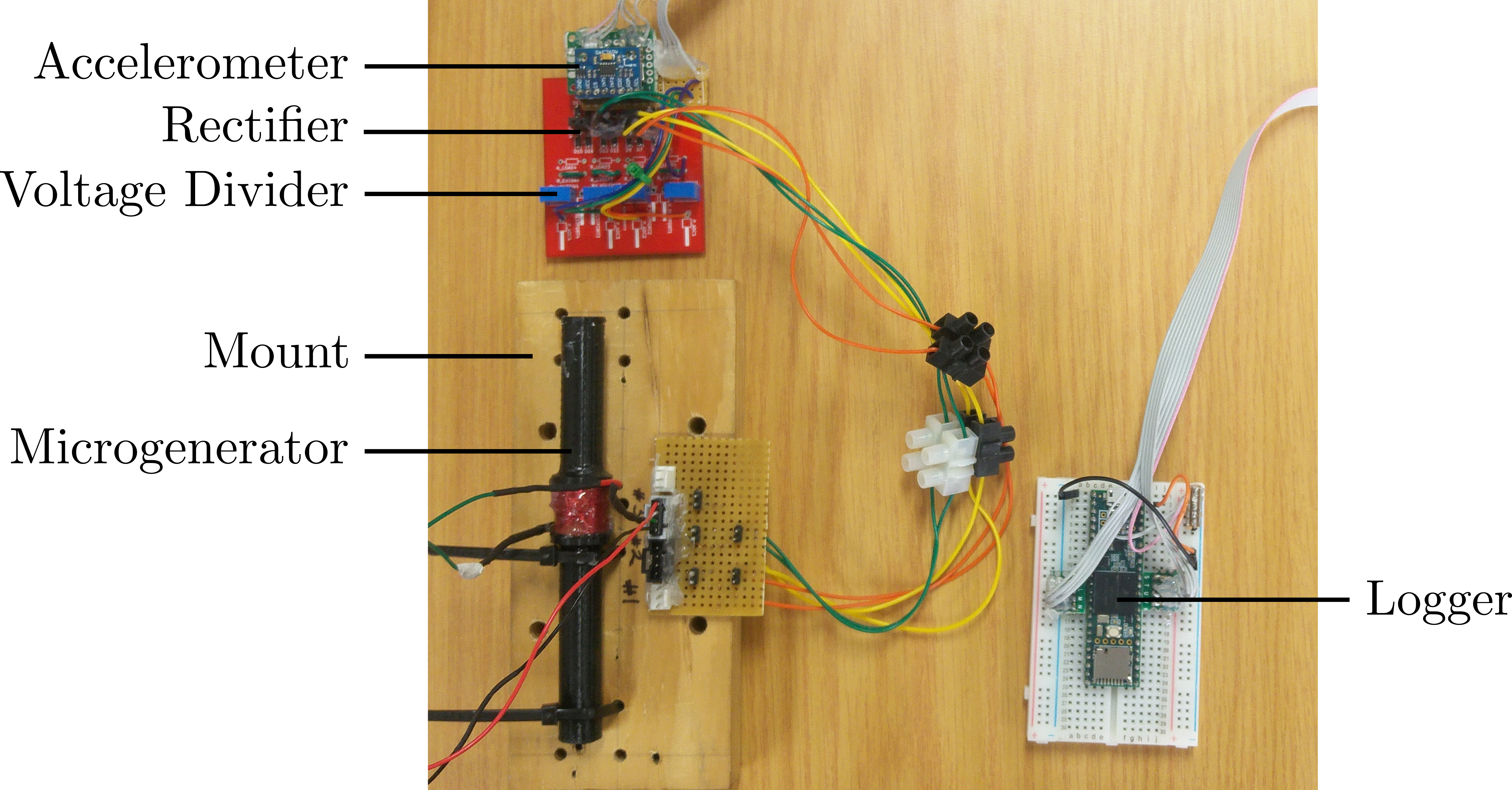}
  \caption{The human control test is performed using a logger, microgenerator
    mount and an electronic circuit consisting of a full-wave bridge rectifier,
    voltage divider and accelerometer.\label{fig:human_test_setup} }
\end{figure}

Two sets of practical tests are performed. The first considers the open-circuit
case, where no load is attached and the open-circuit EMF is measured. The power
that this EMF would deliver to the load can then be calculated using
~\cref{eq:power_avg}. The second test considers the closed-circuit case, where a
load is attached, and the voltage across this load is measured and used to
calculate the power. Hence, the first test evaluates the ideal, open-circuit
model for non-idealized motion, while the second test considers the real-world
practical case. This allows us to assess first the accuracy of our idealized
model, and second our assumption that a microgenerator design that has been
optimized using the idealized model remains optimal when applied to the
closed-circuit case with a real-world load. Our system load is selected as
$R_{\text{load}}=40\Omega$ as discussed in~\cref{sec:elec_sys_model2}.

\subsection{Procedure}
For each test, the subject walked a predetermined short straight course at a normal pace. The
course was approximately 40m in length and perfectly level and clear of obstacles. The course surface
consists of concrete surfaced with laminate vinyl flooring. The course took approximately 35 seconds to
complete. After each test, the microgenerator was alternated to mitigate the
effect of changes in walking speed, style and gait over time. This process is
repeated for a total of 160 tests, 40 times for each microgenerator.

\subsection{Results} \label{sec:results}
\begin{figure}[h!]
  \centering
  \includegraphics[width=\textwidth]{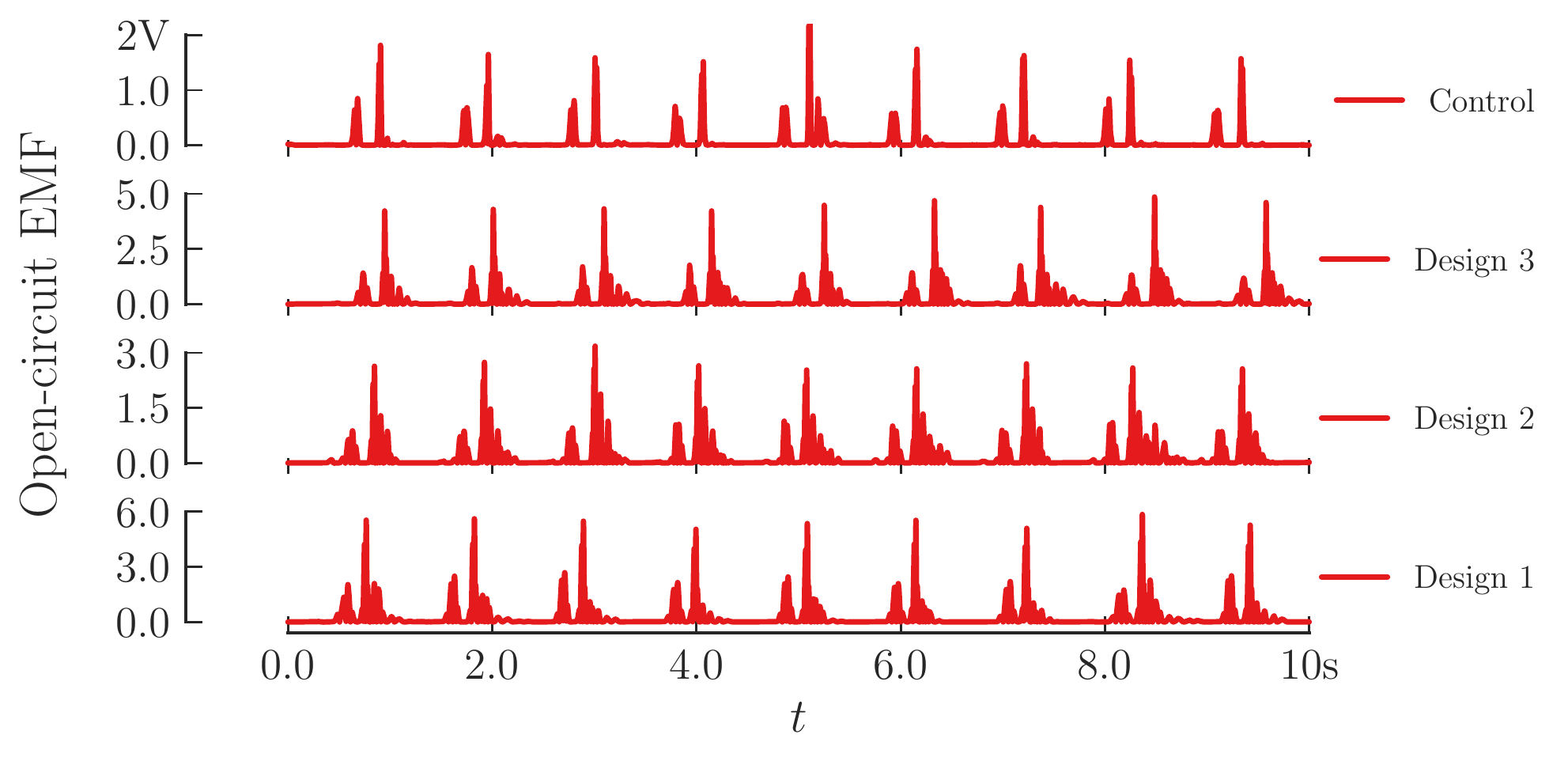}
  \caption{Excerpt from the test measurements of the open-circuit EMF for the
    microgenerator configurations shown in~\cref{fig:tubes_mag_assemblies}. A
    total of 9 footsteps are shown. \label{fig:open_circuit_emf_waveform} }
\end{figure}

A sample of the measured open-circuit EMF for each microgenerator configuration
is shown in \cref{fig:open_circuit_emf_waveform}. Two pulses are seen for
each footstep, the first substantially smaller than the second.  The first occurs during
the deceleration phase and downward acceleration phase of the footstep,
$t_{\text{a}} < t < t_{\text{d}}$ with $t$ as indicated
in~\cref{fig:footstep_extra}. The second occurs after impact at $t \geq
t_{\text{d}}$, and is followed by some residual EMF induced via oscillation of
the magnet assembly on the magnetic spring. 

\begin{figure}[h!]
  \centering
  \includegraphics[width=\textwidth]{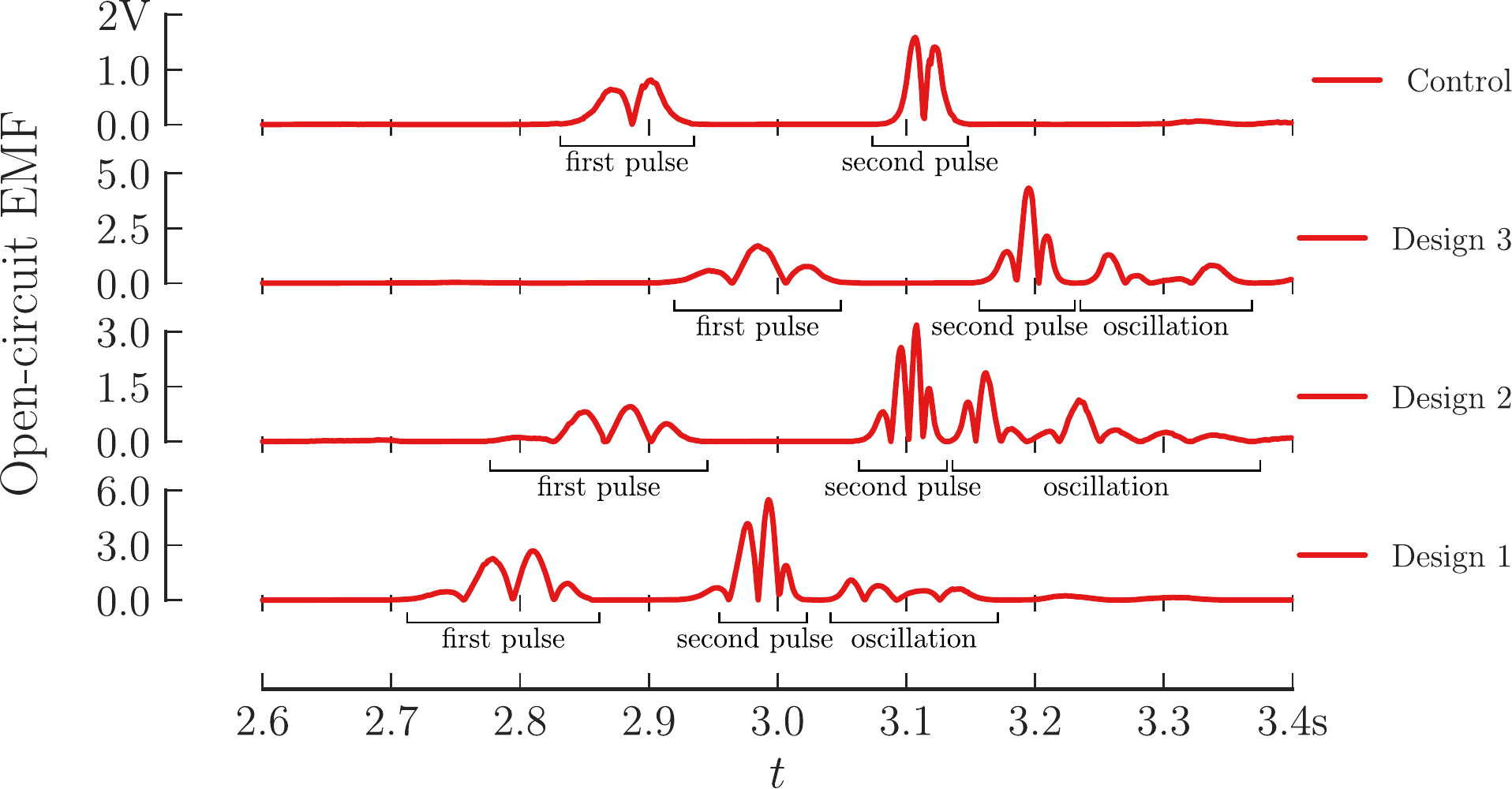}
  \caption{Measured open-circuit EMF for the microgenerator configurations shown
    in \cref{fig:tubes_mag_assemblies} over an interval corresponding to a
    single footstep.\label{fig:open_circuit_emf_segment}. The first EMF pulse
    occurs during the deceleration phase and downward acceleration phase of the
    footstep and the the second pulse occurs after impact, with the footstep
    phases shown in~\cref{fig:footstep_extra}. This is followed by an induced
    EMF due to the oscillation of the magnet assembly.
}
\end{figure}

\Cref{fig:open_circuit_emf_segment} shows the EMF for a single footstep. As
expected, we see that for Design 1 ($c=2, m=2$), Design 2 ($c=1, m=3$) and
Design 3 ($c=1, m=2$) and the Control ($c=1, m=1$) there are four, four, three
and two peaks respectively. Additional minor peaks can be seen following the two
primary pulses as a result of magnet assembly oscillation on the magnetic spring
after passing through the coils. It is interesting to note that more EMF is
induced from this oscillation for Design 2, which also has the longest magnet
assembly ($m=3, s=3\text{mm}$). This allows the upper magnets of the assembly to
more easily reach and induce an EMF in the coils. We do not consider this
oscillation in the proposed model. However, while not very large, it may provide
a means to harvest further energy in future.

The instantaneous power for a series of footsteps for each tested microgenerator
configuration for a $R_{\text{load}}=40 \Omega$ load is calculated from the
open-circuit EMF using~\cref{eq:v_load,eq:power_avg}. 

\begin{figure}[h!]
  \centering \includegraphics[width=\textwidth]{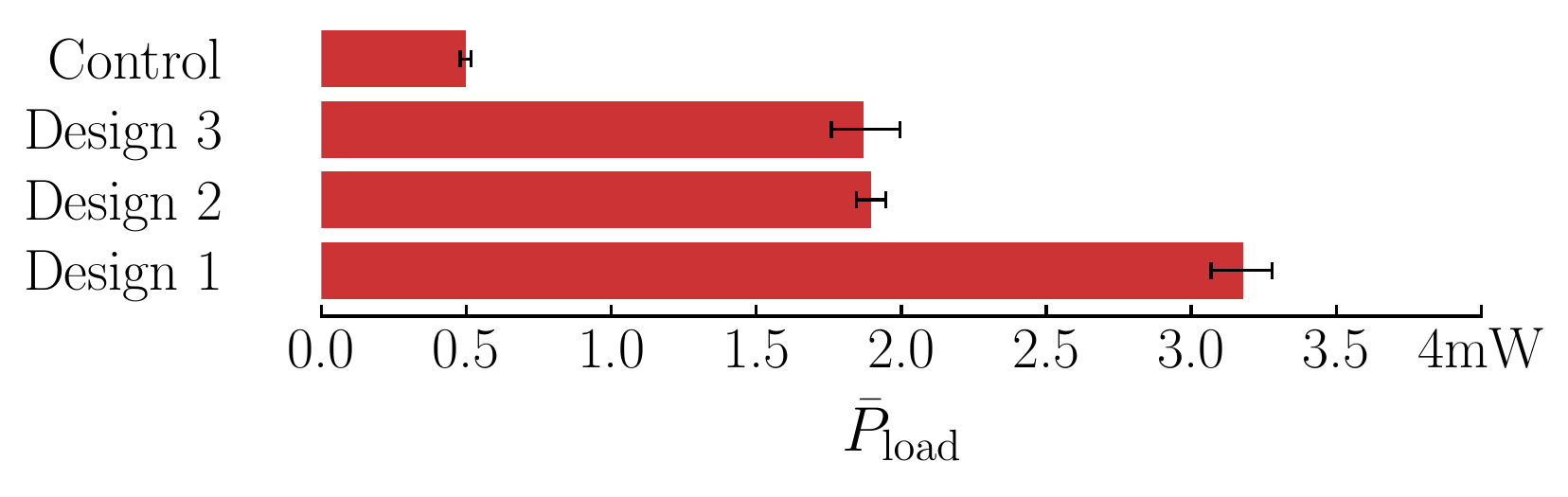}
  \caption{The calculated average power dissipated in a $40\Omega$ load for each
    tested microgenerator configuration given the measured open-circuit RMS of
    the EMF. The Design 1, Design 2, Design 3 and Control configurations have a
    median average power of 3.144mW, 1.886mW, 1.736mW and 0.512mW respectively.
  }
  \label{fig:human_power_measure}
\end{figure}

The distribution of the calculated average power dissipated in the load
$\bar{P}_{\text{load}}$ for each tested microgenerator configuration calculated
using~\cref{eq:power_avg} is illustrated in \cref{fig:human_power_measure}. It
shows that the order of the power generated by the practical microgenerators
agrees with the order of the predicated power output shown
in~\cref{fig:model_sim_comparison}. It is also noteworthy that the relative
differences between the median of the tested configurations mirrors that in
\cref{fig:model_sim_comparison}. This provides supporting evidence for our
hypothesis that optimization for idealized motion serves as a functional
substitute for the optimization for non-idealized motion, as discussed in
\cref{sec:general_power}.

\begin{figure}[h!]
  \centering
  \includegraphics[width=\textwidth]{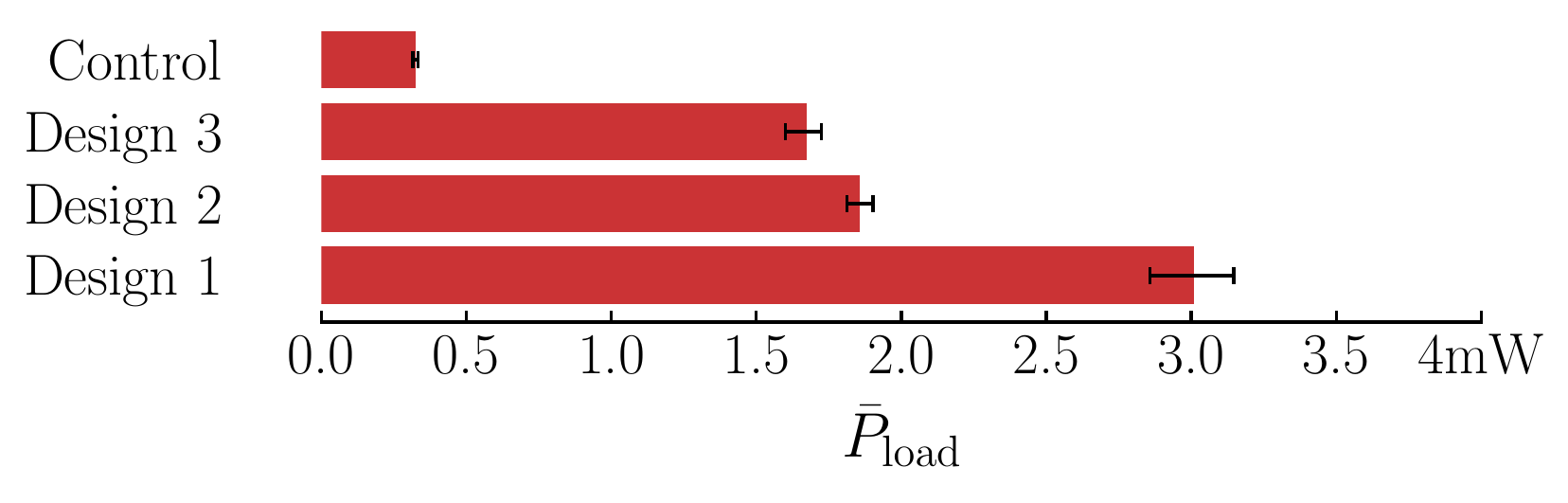}
  \caption{The measured average power dissipated in a $40\Omega$ load for each
    tested microgenerator configuration. Design 1, Design 2, Design 3 and the
    Control configuration deliver median average powers of 3.01mW, 1.856mW,
    1.673mW and 0.324mW respectively.}
  \label{fig:human_power_measure_with_load}
\end{figure}

For the closed-circuit case, the distribution of the average power dissipated in
the load for each tested microgenerator configuration is shown in
\cref{fig:human_power_measure_with_load}. We see that the relative power
delivered to the load by each microgenerator agrees with the idealized power
output that was presented in \cref{fig:model_sim_comparison}. This means that
the best microgenerator design in terms of the idealized model remains the best
design when attached to a practical load, and hence that the idealized model is
suitable for the purpose of optimizing the microgenerator design. Note also that
the power output in \cref{fig:human_power_measure_with_load} is slightly lower
than the corresponding open-circuit power shown in
\cref{fig:human_power_measure}. This is due to the electro-mechanical coupling
that is present in the closed-circuit case.

The results shown in
\cref{fig:human_power_measure,fig:human_power_measure_with_load} indicate strong
supporting evidence for the assumption that the optimization of the
microgenerator design when assuming idealized motion serves as a functional
substitute for optimization under non-idealized motion, as discussed in
\cref{sec:general_power}.

The first design, which delivers median load power of 3.01mW, produces the most
power by a large margin. The second and third designs produce similar levels of
power (1.856mW and 1.673mW) while the control produces substantially less power
with a median of 0.324mW. If we consider the 3.01mW available on average from
Device 1 we note that this is likely to be sufficient to power an energy
harvesting circuit and one of the many ultra-low power microcontroller units
commercially available today. With sufficient power-saving measures, it is quite
possible to envision a self-sustaining system powered by energy harvested from
the sporadic kinetic motion associated with human or animal footsteps.

By calculating the physical volume occupied by each microgenerator we can calculate the
power density of each configuration, measured in $\mu\text{W}/ \text{cm}^3$,
shown in \cref{tab:power_density}. There appears to be a strong correspondence
between the number of coils and magnets and the power density. This is an
expected result given the superposition of subsequent EMF pulses discussed in
\cref{sec:general_power}, which allows significant increases in EMF without
commensurable increases in device length and, hence, volume.

\begin{table}[h!]
  \centering
  \begin{tabular}{@{}lSSS@{}}
    \toprule
    Design & {Median $\bar{P}_{\text{load}}$} & {Volume ($\text{cm}^3$)} & {Power density ($\mu\text{W}/\text{cm}^3$)} \\ \midrule
    1 & 3.010 & 16.78 & 179.380 \\
    2 & 1.856 & 15.11 & 122.833 \\
    3 & 1.673 & 15.72 & 106.425 \\
    4 & 0.324 & 15.46 & 20.957 \\ \bottomrule
  \end{tabular}
  \caption{The median average load power dissipated in the load, microgenerator
    volume and power density of each tested microgenerator configuration.}
  \label{tab:power_density}
\end{table}

%% file: applications.tex
\section{Potential applications}
The method we have proposed is applicable to any form of motion that can be
modelled as a series of impulsive forces along the z-axis, and not just the
footstep-like motion we use to demonstrate effectiveness. Since the formulation
explicitly allows the imposition of constraints such as vertical height and
cross-section limits, it is well-suited for the design of microgenerators that
provide optimal energy generation in size- or weight-constrained situations.
Alternatively, it allows a case-by-case assessment of the feasibility of kinetic
microgeneration based on the design parameters, load and input forces. The
computational model employed eliminates the need for iterative prototyping and
practical testing, thereby reducing both the time and the cost of microgenerator
design.

The particular application which has motivated this research is wildlife
tracking, where size and weight limitations are severe and it is essential to
harvest as much power as possible within these constraints. While this remains
our first intended practical application, we believe there may be many others.
These include energy harvesting from human walking to power wearable technology,
and from the impulse-like nature of uneven road surfaces to power autonomous
vehicle-mounted sensors.

%% file: conclusion.tex
\section{Conclusion}
We have considered a linear kinetic energy harvester architecture that consists
of an assembly of one or more spaced magnets that passes through one or more
coils when the device experiences motion along its axis and that is suspended by a magnetic
spring. We considered the specific case of impulsive acceleration as might be
the results of a footstep, and not harmonic vibration usually assumed for
kinetic energy harvesting. We introduced a mechanical and electrical system
model that allows this microgenerator architecture to be optimized for power
supplied to a load in terms of its design parameters. These parameters include
the individual coil height, number of coils, number of magnets in the magnet
assembly and the relative spacing between the coils and magnets. By deliberately
designing the model to allow the incorporation of constraints and by selecting a
compatible optimization technique, we are able to adapt our architecture
to any impulse-like excitation provided there is sufficient single-axis motion.

Our technique was evaluated by application to the practical scenario of
designing a microgenerator that can be worn on the leg of a human or animal.
First, we predict the top three designs given the physical size constraints.
Next, we build physical prototypes of these three designs and measure their
performance when attached to the leg of a human subject while walking. In all
cases, a baseline system with the simplest possible design is also evaluated. We
find that the theoretically predicted relative ordering of the produced power
agrees with that observed for the practical systems. This demonstrates that the
theoretical optimization also led to practically optimal results. In all cases,
the designed system far outperformed the baseline.

The best microgenerator configuration achieved an average load power of 3.144mW,
supplied to a $40\Omega$ load giving a power density of 179.380$\mu W /
\text{cm}^3$ from foot impact accelerations of approximately 2.2g. This firmly
places the best micrognerator configuration in the realm of ultra-low power
microcontrollers, making powering of such devices a realistic possibility.


%% file: acknowledgements.tex
\section{Acknowledgments}
The authors gratefully acknowledge financial support by the National Research
Foundation of the Republic of South Africa, by Telkom South Africa, and by
Innovus of Stellenbosch University. The authors also gratefully acknowledge the
invaluable assistance of Mr Wessel Croukamp in the physical construction of the
devices.

%% file: funding.tex
\section{Funding declaration}
We wish to confirm and declare the following sources of funding and/or research
grants that were received in the course of study, research or assembly of this
work:

\begin{itemize}
\item The National Research Foundation (NRF) of South Africa
\item Telkom South Africa
  \item Innovus of Stellenbosch University
  \end{itemize}

The above sponsors have not played any role in the study design; collection;
analysis and interpretation of data in the writing of the report; and in the
decision to submit the article for publication.


%% file: main.bbl
\begin{thebibliography}{10}
\providecommand{\url}[1]{#1}
\csname url@samestyle\endcsname
\providecommand{\newblock}{\relax}
\providecommand{\bibinfo}[2]{#2}
\providecommand{\BIBentrySTDinterwordspacing}{\spaceskip=0pt\relax}
\providecommand{\BIBentryALTinterwordstretchfactor}{4}
\providecommand{\BIBentryALTinterwordspacing}{\spaceskip=\fontdimen2\font plus
\BIBentryALTinterwordstretchfactor\fontdimen3\font minus
  \fontdimen4\font\relax}
\providecommand{\BIBforeignlanguage}[2]{{%
\expandafter\ifx\csname l@#1\endcsname\relax
\typeout{** WARNING: IEEEtran.bst: No hyphenation pattern has been}%
\typeout{** loaded for the language `#1'. Using the pattern for}%
\typeout{** the default language instead.}%
\else
\language=\csname l@#1\endcsname
\fi
#2}}
\providecommand{\BIBdecl}{\relax}
\BIBdecl

\bibitem{Conrad2008}
\BIBentryALTinterwordspacing
J.~M. Conrad, ``{A survey of energy harvesting sources for embedded systems},''
  in \emph{IEEE SoutheastCon 2008}.\hskip 1em plus 0.5em minus 0.4em\relax
  IEEE, apr 2008, pp. 442--447. [Online]. Available:
  \url{http://ieeexplore.ieee.org/lpdocs/epic03/wrapper.htm?arnumber=4494336}
\BIBentrySTDinterwordspacing

\bibitem{Vullers2009}
\BIBentryALTinterwordspacing
R.~Vullers, R.~van Schaijk, I.~Doms, C.~{Van Hoof}, and R.~Mertens,
  ``{Micropower energy harvesting},'' \emph{Solid-State Electronics}, vol.~53,
  no.~7, pp. 684--693, jul 2009. [Online]. Available:
  \url{http://www.sciencedirect.com/science/article/pii/S0038110109000720}
\BIBentrySTDinterwordspacing

\bibitem{Sudevalayam2011}
\BIBentryALTinterwordspacing
S.~Sudevalayam and P.~Kulkarni, ``{Energy Harvesting Sensor Nodes: Survey and
  Implications},'' \emph{IEEE Communications Surveys {\&} Tutorials}, vol.~13,
  no.~3, pp. 443--461, 2011. [Online]. Available:
  \url{http://ieeexplore.ieee.org/lpdocs/epic03/wrapper.htm?arnumber=5522465}
\BIBentrySTDinterwordspacing

\bibitem{Gilbert2008}
\BIBentryALTinterwordspacing
J.~M. Gilbert and F.~Balouchi, ``{Comparison of energy harvesting systems for
  wireless sensor networks},'' \emph{International Journal of Automation and
  Computing}, vol.~5, no.~4, pp. 334--347, oct 2008. [Online]. Available:
  \url{http://link.springer.com/10.1007/s11633-008-0334-2}
\BIBentrySTDinterwordspacing

\bibitem{Bobryk2016}
\BIBentryALTinterwordspacing
R.~V. Bobryk and D.~Yurchenko, ``{On enhancement of vibration-based energy
  harvesting by a random parametric excitation},'' \emph{Journal of Sound and
  Vibration}, vol. 366, pp. 407--417, 2016. [Online]. Available:
  \url{http://dx.doi.org/10.1016/j.jsv.2015.11.033}
\BIBentrySTDinterwordspacing

\bibitem{Wang2017}
\BIBentryALTinterwordspacing
W.~Wang, J.~Cao, N.~Zhang, J.~Lin, and W.~H. Liao, ``{Magnetic-spring based
  energy harvesting from human motions: Design, modeling and experiments},''
  \emph{Energy Conversion and Management}, vol. 132, pp. 189--197, 2017.
  [Online]. Available: \url{http://dx.doi.org/10.1016/j.enconman.2016.11.026}
\BIBentrySTDinterwordspacing

\bibitem{Yang2017}
\BIBentryALTinterwordspacing
W.~Yang and S.~Towfighian, ``{A hybrid nonlinear vibration energy harvester},''
  \emph{Mechanical Systems and Signal Processing}, vol.~90, pp. 317--333, 2017.
  [Online]. Available: \url{http://dx.doi.org/10.1016/j.ymssp.2016.12.032}
\BIBentrySTDinterwordspacing

\bibitem{Haroun2015}
\BIBentryALTinterwordspacing
A.~Haroun, I.~Yamada, and S.~Warisawa, ``{Study of electromagnetic vibration
  energy harvesting with free/impact motion for low frequency operation},''
  \emph{Journal of Sound and Vibration}, vol. 349, pp. 389--402, 2015.
  [Online]. Available:
  \url{http://www.sciencedirect.com/science/article/pii/S0022460X15002874}
\BIBentrySTDinterwordspacing

\bibitem{Kecik2017}
\BIBentryALTinterwordspacing
K.~Kecik, A.~Mitura, and J.~Warminski, ``{Energy harvesting from a magnetic
  levitation system},'' \emph{International Journal of Non-Linear Mechanics},
  vol.~94, pp. 200--206, sep 2017. [Online]. Available:
  \url{https://www.sciencedirect.com/science/article/pii/S002074621730241X?via{\%}3Dihub}
\BIBentrySTDinterwordspacing

\bibitem{Marszal2017}
M.~Marszal, B.~Witkowski, K.~Jankowski, P.~Perlikowski, and T.~Kapitaniak,
  ``{Energy harvesting from pendulum oscillations},'' \emph{International
  Journal of Non-Linear Mechanics}, vol.~94, no. April, pp. 251--256, 2017.

\bibitem{SoaresdosSantos2016}
\BIBentryALTinterwordspacing
M.~P. {Soares dos Santos}, J.~A.~F. Ferreira, J.~A.~O. Sim{\~{o}}es,
  R.~Pascoal, J.~Torr{\~{a}}o, X.~Xue, and E.~P. Furlani, ``{Magnetic
  levitation-based electromagnetic energy harvesting: a semi-analytical
  non-linear model for energy transduction},'' \emph{Scientific Reports},
  vol.~6, no.~1, p. 18579, may 2016. [Online]. Available:
  \url{http://www.nature.com/articles/srep18579}
\BIBentrySTDinterwordspacing

\bibitem{Berdy2015}
\BIBentryALTinterwordspacing
D.~F. Berdy, D.~J. Valentino, and D.~Peroulis, ``{Kinetic energy harvesting
  from human walking and running using a magnetic levitation energy
  harvester},'' \emph{Sensors and Actuators, A: Physical}, vol. 222, pp.
  262--271, 2015. [Online]. Available:
  \url{http://dx.doi.org/10.1016/j.sna.2014.12.006}
\BIBentrySTDinterwordspacing

\bibitem{Serre2008}
\BIBentryALTinterwordspacing
C.~Serre, A.~P{\'{e}}rez-Rodr{\'{i}}guez, N.~Fondevilla, E.~Martincic,
  S.~Mart{\'{i}}nez, J.~R. Morante, J.~Montserrat, and J.~Esteve, ``{Design and
  implementation of mechanical resonators for optimized inertial
  electromagnetic microgenerators},'' \emph{Microsystem Technologies}, vol.~14,
  no. 4-5, pp. 653--658, apr 2008. [Online]. Available:
  \url{http://link.springer.com/10.1007/s00542-007-0494-y}
\BIBentrySTDinterwordspacing

\bibitem{Kwon2013}
\BIBentryALTinterwordspacing
S.-D. Kwon, J.~Park, and K.~Law, ``{Electromagnetic energy harvester with
  repulsively stacked multilayer magnets for low frequency vibrations},''
  \emph{Smart Materials and Structures}, vol.~22, no.~5, p. 055007, may 2013.
  [Online]. Available:
  \url{http://stacks.iop.org/0964-1726/22/i=5/a=055007?key=crossref.a2cc56191125641b68f3d636a2068367}
\BIBentrySTDinterwordspacing

\bibitem{Wickenheiser2010}
\BIBentryALTinterwordspacing
A.~M. Wickenheiser and E.~Garcia, ``{Broadband vibration-based energy
  harvesting improvement through frequency up-conversion by magnetic
  excitation},'' \emph{Smart Materials and Structures}, vol.~19, no.~6, p.
  065020, jun 2010. [Online]. Available:
  \url{http://stacks.iop.org/0964-1726/19/i=6/a=065020?key=crossref.7c6533acecb40702c342818b6a2f79ee}
\BIBentrySTDinterwordspacing

\bibitem{Ylli2015}
\BIBentryALTinterwordspacing
K.~Ylli, D.~Hoffmann, A.~Willmann, P.~Becker, B.~Folkmer, and Y.~Manoli,
  ``{Energy harvesting from human motion: exploiting swing and shock
  excitations},'' \emph{Smart Materials and Structures}, vol.~24, no.~2, p.
  025029, feb 2015. [Online]. Available:
  \url{http://stacks.iop.org/0964-1726/24/i=2/a=025029?key=crossref.7f9622e87f20f174a5b5a2dd614d71c7}
\BIBentrySTDinterwordspacing

\bibitem{Saravia2017}
\BIBentryALTinterwordspacing
C.~M. Saravia, J.~M. Ram{\'{i}}rez, and C.~D. Gatti, ``{A hybrid
  numerical-analytical approach for modeling levitation based vibration energy
  harvesters},'' \emph{Sensors and Actuators, A: Physical}, vol. 257, pp.
  20--29, 2017. [Online]. Available:
  \url{http://dx.doi.org/10.1016/j.sna.2017.01.023}
\BIBentrySTDinterwordspacing

\bibitem{Carroll2012}
\BIBentryALTinterwordspacing
D.~Carroll and M.~Duffy, ``{Modelling, design, and testing of an
  electromagnetic power generator optimized for integration into shoes},''
  \emph{Proceedings of the Institution of Mechanical Engineers, Part I: Journal
  of Systems and Control Engineering}, vol. 226, no.~2, pp. 256--270, feb 2012.
  [Online]. Available:
  \url{http://pii.sagepub.com/lookup/doi/10.1177/0959651811411406}
\BIBentrySTDinterwordspacing

\bibitem{leRoux2017}
\BIBentryALTinterwordspacing
S.~P. le~Roux, J.~Marias, R.~Wolhuter, and T.~Niesler, ``{Animal-borne
  behaviour classification for sheep (Dohne Merino) and Rhinoceros
  (Ceratotherium simum and Diceros bicornis)},'' \emph{Animal Biotelemetry},
  vol.~5, no.~1, p.~25, dec 2017. [Online]. Available:
  \url{https://animalbiotelemetry.biomedcentral.com/articles/10.1186/s40317-017-0140-0}
\BIBentrySTDinterwordspacing

\bibitem{Blackie2010}
\BIBentryALTinterwordspacing
H.~M. Blackie, ``{Comparative Performance of Three Brands of Lightweight Global
  Positioning System Collars},'' \emph{Journal of Wildlife Management},
  vol.~74, no.~8, pp. 1911--1916, nov 2010. [Online]. Available:
  \url{http://www.bioone.org/doi/abs/10.2193/2009-412}
\BIBentrySTDinterwordspacing

\bibitem{hebblewhite2007}
\BIBentryALTinterwordspacing
M.~Hebblewhite, M.~Percy, and E.~H. Merrill, ``{Are All Global Positioning
  System Collars Created Equal? Correcting Habitat-Induced Bias Using Three
  Brands in the Central Canadian Rockies},'' \emph{Journal of Wildlife
  Management}, vol.~71, no.~6, pp. 2026--2033, aug 2007. [Online]. Available:
  \url{http://www.bioone.org/doi/abs/10.2193/2006-238}
\BIBentrySTDinterwordspacing

\bibitem{Masoumi2016}
\BIBentryALTinterwordspacing
M.~Masoumi and Y.~Wang, ``{Repulsive magnetic levitation-based ocean wave
  energy harvester with variable resonance: Modeling, simulation and
  experiment},'' \emph{Journal of Sound and Vibration}, vol. 381, pp. 192--205,
  2016. [Online]. Available: \url{http://dx.doi.org/10.1016/j.jsv.2016.06.024}
\BIBentrySTDinterwordspacing

\bibitem{Donoso2009}
\BIBentryALTinterwordspacing
G.~Donoso, C.~L. Ladera, and P.~Mart{\'{i}}n, ``{Magnet fall inside a
  conductive pipe: motion and the role of the pipe wall thickness},''
  \emph{Eur. J. Phys. J. Phys}, vol.~30, no.~30, pp. 855--855, 2009. [Online].
  Available: \url{http://iopscience.iop.org/0143-0807/30/4/018}
\BIBentrySTDinterwordspacing

\bibitem{VonBuren2007}
\BIBentryALTinterwordspacing
T.~von B{\"{u}}ren and G.~Tr{\"{o}}ster, ``{Design and optimization of a linear
  vibration-driven electromagnetic micro-power generator},'' \emph{Sensors and
  Actuators A: Physical}, vol. 135, no.~2, pp. 765--775, 2007. [Online].
  Available:
  \url{http://www.sciencedirect.com.ez.sun.ac.za/science/article/pii/S0924424706005413}
\BIBentrySTDinterwordspacing

\bibitem{Khan2014}
\BIBentryALTinterwordspacing
F.~Khan, B.~Stoeber, F.~Sassani, F.~Khan, B.~Stoeber, and F.~Sassani,
  ``{Modeling and simulation of linear and nonlinear MEMS scale electromagnetic
  energy harvesters for random vibration environments.}''
  \emph{TheScientificWorldJournal}, vol. 2014, p. 742580, 2014. [Online].
  Available: \url{http://www.ncbi.nlm.nih.gov/pubmed/24605063}
\BIBentrySTDinterwordspacing

\bibitem{Zeng2013}
\BIBentryALTinterwordspacing
P.~Zeng and A.~Khaligh, ``{A Permanent-Magnet Linear Motion Driven Kinetic
  Energy Harvester},'' \emph{IEEE Transactions on Industrial Electronics},
  vol.~60, no.~12, pp. 5737--5746, dec 2013. [Online]. Available:
  \url{http://ieeexplore.ieee.org/document/6359913/}
\BIBentrySTDinterwordspacing

\bibitem{Vokoun2009}
D.~Vokoun, M.~Beleggia, L.~Heller, and P.~{\v{S}}ittner, ``{Magnetostatic
  interactions and forces between cylindrical permanent magnets},''
  \emph{Journal of Magnetism and Magnetic Materials}, vol. 321, no.~22, pp.
  3758--3763, 2009.

\bibitem{Mann2009}
B.~P. Mann and N.~D. Sims, ``{Energy harvesting from the nonlinear oscillations
  of magnetic levitation},'' \emph{Journal of Sound and Vibration}, vol. 319,
  no. 1-2, pp. 515--530, 2009.

\bibitem{Kraft1988}
D.~Kraft, ``{A software package for sequential quadratic programming},''
  Institute for Flight Mechanics, Koln, Germany, Tech. Rep., 1988.

\bibitem{Bedekar2009}
\BIBentryALTinterwordspacing
V.~Bedekar, J.~Oliver, and S.~Priya, ``{Pen harvester for powering a pulse rate
  sensor},'' \emph{Journal of Physics D: Applied Physics}, vol.~42, no.~10, p.
  105105, may 2009. [Online]. Available:
  \url{http://stacks.iop.org/0022-3727/42/i=10/a=105105?key=crossref.6e3dc42c217b694ce465a1ab208f2f8a}
\BIBentrySTDinterwordspacing

\end{thebibliography}
